\documentclass[aps,prb,preprint,superscriptaddress,showpacs]{revtex4-1}
\usepackage{graphicx}
\usepackage{hyperref}
\hypersetup{
    colorlinks=false,
    citebordercolor=[rgb]{0,0,1}
}
\usepackage{array}
\newcolumntype{C}[1]{>{\centering\let\newline\\\arraybackslash\hspace{0pt}}m{#1}}
\begin{document}

%\preprint{}

\title{Pressure-induced spin pairing transition of Fe$^{3+}$ in oxygen octahedra}


\author{D. M. Vasiukov}
\email[]{vasyukov@physics.msu.ru}
%\thanks{}
\affiliation{Laboratory of Crystallography, Universit\"{a}t Bayreuth, D-95440 Bayreuth, Germany}
\affiliation{Bayerisches Geoinstitut, Universit\"{a}t Bayreuth, D-95440 Bayreuth, Germany}
\author{L. Dubrovinsky}
\affiliation{Bayerisches Geoinstitut, Universit\"{a}t Bayreuth, D-95440 Bayreuth, Germany}
\author{I. Kupenko}
\affiliation{Institut f\"{u}r Mineralogie, Universit\"{a}t M\"{u}nster, D-48149 M\"{u}nster, Germany}
\author{V. Cerantola}
\affiliation{ESRF-The European Synchrotron CS40220 38043 Grenoble Cedex 9 France}
\author{G. Aprilis}
\affiliation{Laboratory of Crystallography, Universit\"{a}t Bayreuth, D-95440 Bayreuth, Germany}
\affiliation{Bayerisches Geoinstitut, Universit\"{a}t Bayreuth, D-95440 Bayreuth, Germany}
\author{L. Ismailova}
\affiliation{Skolkovo Institute of Science and Technology, Skolkovo Innovation Center, 143026 Moscow, Russia}
\author{E. Bykova}
\affiliation{Photon Science, Deutsches Elektronen-Synchrotron, D-22607 Hamburg, Germany}
\author{C. McCammon}
\affiliation{Bayerisches Geoinstitut, Universit\"{a}t Bayreuth, D-95440 Bayreuth, Germany}
\author{C. Prescher}
\affiliation{Institut f\"{u}r Geologie und Mineralogie, Universit\"{a}t zu K\"{o}ln, D-50674 K\"{o}ln, Germany}
\author{A. I. Chumakov}
\affiliation{ESRF-The European Synchrotron CS40220 38043 Grenoble Cedex 9 France}
\author{N. Dubrovinskaia}
\affiliation{Laboratory of Crystallography, Universit\"{a}t Bayreuth, D-95440 Bayreuth, Germany}


\date{\today}

\begin{abstract}
High pressure can provoke spin transitions in transition metal-bearing compounds. These transitions are of high interest not only for fundamental physics and chemistry, but also may have important implications for geochemistry and geophysics of the Earth and planetary interiors. Here we have carried out a comparative study of the pressure-induced spin transition in compounds with trivalent iron, octahedrally coordinated by oxygen. High-pressure single-crystal M\"{o}ssbauer spectroscopy data for FeBO$_3$, Fe$_2$O$_3$ and Fe$_3$(Fe$_{1.766(2)}$Si$_{0.234(2)}$)(SiO$_4$)$_3$ are presented together with detailed analysis of hyperfine parameter behavior. We argue that $\zeta$-Fe$_2$O$_3$ is an intermediate phase in the reconstructive phase transition between $\iota$-Fe$_2$O$_3$ and $\theta$-Fe$_2$O$_3$ and question the proposed perovskite-type structure for $\zeta$-Fe$_2$O$_3$.The structural data show that the spin transition is closely related to the volume of the iron octahedron. The transition starts when volumes reach 8.9--9.3~\AA$^3$, which corresponds to pressures of 45--60~GPa, depending on the compound. Based on phenomenological arguments we conclude that the spin transition can proceed only as a first-order phase transition in magnetically-ordered compounds. An empirical rule for prediction of cooperative behavior at the spin transition is proposed. The instability of iron octahedra, together with strong interactions between them in the vicinity of the critical volume, may trigger a phase transition in the metastable phase. We find that the isomer shift of high spin iron ions depends linearly on the octahedron volume with approximately the same coefficient, independent of the particular compounds and/or oxidation state. For eight-fold coordinated Fe$^{2+}$ we observe a significantly weaker nonlinear volume dependence.
\end{abstract}

%\keywords{}

\maketitle

\section{\label{introduction}Introduction}

The spin-pairing transition in transition metal compounds, also known as spin crossover or spin transition (the first observation published in 1931~\cite{cambi1931magnetische}), is an important phenomenon that can cause drastic changes in physical properties of materials, including alterations in volume, compressibility and electrical conductivity. Upon crossover from high spin (HS) to low spin (LS), the ionic radii change dramatically  (for example, for Fe$^{2+}$ from 78 to 61~pm, and Fe$^{3+}$ from 64.5 to 55~pm~\cite{shannon1976revised}) that leads to a reduction of chemical bond lengths and may cause structural changes. Reduction of the radii means that the LS ion can substitute for other cations without significant change of the crystal structure, which is important for chemistry and especially for geochemistry as it may lead to variations in partitioning of elements (particularly iron, one of the most abundant elements in the Earth and a component of major mantle-forming minerals). Magnetic properties are also affected: for example, octahedrally-coordinated ferrous iron (Fe$^{2+}$) in the LS state is diamagnetic. Such versatile behavior has motivated ongoing attempts to find practical applications for this phenomenon, for instance, as storage media and in displays~\cite{kahn1998spin}, temperature-sensitive contrast agents for magnetic resonance imaging~\cite{muller2003spin}, and as a mechanical actuator~\cite{shepherd2013molecular} (see review~\onlinecite{molnar2014emerging}).

\emph{Thermally-induced} spin crossover is a frequent phenomenon in coordination complexes with suitable ligands when the crystal-field splitting parameter $D_q$ is close to the mean spin-pairing energy $E_p$. Most of such compounds are complexes of iron (see the recent review~\onlinecite{gutlich2013spin} on this topic).

A \emph{pressure-induced} spin transition can be observed even in compounds with low ambient $D_q$, because the crystal-field splitting parameter increases upon compression. However, until recently such investigations were rare. The swift development of the diamond anvil cell (DAC) technique yielded several striking discoveries, such as, for example, spin crossover in ferropericlase at lower mantle conditions~\cite{badro2003iron}. M\"{o}ssbauer spectroscopy (MS) coupled with single crystal X-ray diffraction (XRD) is an extremely powerful combination for investigation of spin transitions. High-pressure crystallography using synchrotron facilities with focused high-energy X-rays and fast 2D detectors, and use of DACs with a large opening angle~\cite{boehler2004new} enable careful investigation of the geometry of the FeO$_6$ octahedron before and after the spin transition and provide a basis to search for correlations between crystal chemistry and hyperfine parameters of the compounds under investigation.

In this paper we focus on ferric iron (Fe$^{3+}$) octahedrally coordinated by oxygen for the following reasons: (\emph{i}) the Fe$^{3+}$O$_6$ octahedron is a common structural element in different compounds and minerals (iron-containing garnets, perovskite-structured materials including ferrites,  simple (Fe$_2$O$_3$) and complex oxides, etc.), (\emph{ii}) this polyhedron is of interest for physics, as a spin transition in an ion with $d^5$-configuration may provoke an insulator-metal transition~\cite{ovchinnikov2008effect}, (\emph{iii}) oxygen has the second highest value of electronegativity after fluorine, so that the Fe-O bond shows a marked ionic character, and (\emph{iv}) high quality experimental data on spin transitions in some Fe$^{3+}$O$_6$-containing compounds are already available. Complemented by our new experimental results, they may be useful for a comparative analysis aimed at establishing regularities in the pressure-induced spin transition of trivalent iron in the oxygen octahedron. The following compounds were examined: iron borate (FeBO$_3$), hematite (Fe$_2$O$_3$), skiagite-rich garnet (Fe$_5$(SiO$_4$)$_3$), goethite (FeOOH), calcium ferrite (CaFe$_2$O$_4$) and andradite (Ca$_3$Fe$_2$(SiO$_4$)$_3$) (Table~\ref{Used data}).

\begin{table*}[b]
\caption{\label{Used data}List of compounds examined in the present work. References to literature data used in the comparative analysis are provided. Hereafter an em dash means the absence of data.}
\begin{ruledtabular}
\begin{tabular}{cC{3cm}C{8cm}}
Chemical compound & Crystallographic data & M\"{o}ssbauer spectroscopy data kind of material/reference\\
\hline
FeBO$_3$ &  Ref.~\onlinecite{Greenberg_be_published} & powder, conventional MS, and single-crystal, MS with SMS/this study  \\
Fe$_2$O$_3$ &  Ref.~\onlinecite{schouwink2011high} & single-crystal, MS with SMS/ this study \\
Fe$_3$(Fe$_{1.766(2)}$Si$_{0.234(2)}$)(SiO$_4$)$_3$ & Ref.~\onlinecite{ismailova2017effect} & single-crystal, MS with SMS/ this study\\
FeOOH & Ref.~\onlinecite{xu2013pressure} & powder/Ref.~\onlinecite{xu2013pressure} \\
CaFe$_2$O$_4$ & Ref.~\onlinecite{merlini2010letter} & powder/Ref.~\onlinecite{greenberg2013mott} \\
Ca$_3$Fe$_2$Si$_3$O$_{12}$ & Ref.~\onlinecite{friedrich2014pressure} & ---\\
\end{tabular}
\end{ruledtabular}
\end{table*}

FeBO$_3$ has a calcite (CaCO$_3$)-type structure ($R\bar{3}c$ space group) \cite{bernal1963new,diehl1975crystal}. At ambient conditions the structure consists of slightly distorted corner-shared oxygen octahedra enclosing ferric iron and BO$_3$ triangles oriented in the structure perpendicular to the 3-fold axis. Iron borate is antiferromagnetic with weak ferromagnetism due to small canting of iron spins from antiparallel alignment~\cite{petrov1972nuclear} with N{\'e}el point $T_N=348$~K~\cite{wolfe1970room,eibschutz1970critical}. A nuclear forward scattering (NFS) study showed that FeBO$_3$ undergoes a phase transition to a nonmagnetic state at 46~GPa~\cite{troyan2001transition}.  Single-crystal MS data~\cite{sarkisyan2002magnetic} led to conclusion that it is due to a HS$\rightarrow$LS transition. A recent detailed high-pressure single-crystal XRD study~\cite{Greenberg_be_published} confirmed earlier suppositions~\cite{gavriliuk2002equation,parlinski2002structural} that  it is an isosymmetric transition\footnote{At isosymmetric transition the space group preserves and the atoms do not change their Wyckoff positions. In the literature it is also often called isostructural.}. Thus, FeBO$_3$ is an excellent reference compound for investigation of relations between crystal chemistry and hyperfine parameters of trivalent iron as a function of pressure and geometry of the Fe$^{3+}$O$_6$ octahedron.

Hematite is a well-known iron sesquioxide (Fe$_2$O$_3$) with a corundum-type (Al$_2$O$_3$) structure consisting only of Fe$^{3+}$O$_6$ octahedra. Each octahedron connects with three neighbors via edges in honeycomb-like layers and the layers are interconnected through common triangular faces of octahedra~\cite{blake1966refinement}  (space group $R\bar{3}c$). It is antiferromagnetic (canted-type at ambient conditions) with a high N{\'e}el temperature of 948 K. Upon compression at ambient temperature up to about 100 GPa, it undergoes three phase transformations: pressure-induced Morin transition with reorientation of iron magnetic moments and loss of canting at 1.7~GPa~\cite{klotz2013pressure}, structural transition to the $\zeta$-Fe$_2$O$_3$ phase with a triclinic, distorted-perovskite structure (space group $P$\={1}) at 54 GPa~\cite{bykova2013novel,bykova2016structural}, and a transition to the orthorhombic $\theta$-Fe$_2$O$_3$ phase (space group $Aba2$) at 67~GPa~\cite{bykova2016structural}. The perovskite-type phase consists of two types of iron coordination polyhedra --- octahedra and bicapped triangular prisms. The structural details of this phase are still not fully clarified because of its low (triclinic) symmetry and difficulties in collecting a sufficiently full XRD dataset in the DAC~\cite{bykova2016structural}. The orthorhombic phase contains ferric iron in distorted triangular prisms only and Fe$^{3+}$ is in a LS state, as suggested by available XRD and MS data~\cite{pasternak1999breakdown,bykova2016structural}. There has been only limited information about the behavior of trivalent iron at the spin transition in Fe$_2$O$_3$ polymorphs, and we provide here new experimental data.

Skiagite is an iron end-member in the silicate garnet family with ideal chemical formula\linebreak Fe$^{2+}_3$Fe$^{3+}_2$(SiO$_4$)$_3$. Samples used in the present study are a solid solution of skiagite and iron-majorite (Fe$_3$(FeSi)(SiO$_4$)$_3$), which have been described by Ismailova et al~\cite{ismailova2015high}. Their exact chemical composition Fe$_3$(Fe$_{1.532(2)}^{3+}$Fe$_{0.234(2)}^{2+}$Si$_{0.234(2)}^{4+}$)(SiO$_4$)$_3$ was determined from single-crystal XRD and microprobe analysis~\cite{ismailova2015high}. It corresponds to approximately 23~mol~\% of iron majorite component in the samples. Skiagite-majorite solid solution has a cubic crystal structure (space group $Ia\bar{3}d$). The cubic X-site is populated by ferrous iron, whereas Fe$^{3+}$ and (Fe$^{2+}$Si$^{4+}$) share the octahedral position (Y-site). XRD data revealed isosymmetric crossover accompanied by a drop of the octahedron volume above 50 GPa~\cite{ismailova2017effect}. We collected M\"{o}ssbauer spectra of the skiagite-majorite sample on compression up to 90 GPa and found that the spectra change in the same pressure range where XRD data indicate the phase transition. Although iron occupies two structural positions and has multiple electronic states, our new data allowed unambiguous interpretation of the mechanism and origin of this transition.

Goethite ($\alpha$-FeOOH) is a widespread mineral with a diaspore-type ($\alpha$-AlOOH) structure (space group \emph{Pnma}~\cite{yang2006goethite}).  Iron
octahedra share edges to form double chains along the $c$-axis (in \emph{Pbnm} setting), which are further linked in a three-dimensional structure by sharing vertices. At ambient conditions goethite is a collinear antiferromagnetic~\cite{forsyth1968magnetic,szytula1968neutron} with N{\'e}el temperature $T_N=393$~K~\cite{ozdemir1996thermoremanence}. A sharp spin transition accompanied by hydrogen bond symmetrization was revealed at 45~GPa~\cite{xu2013pressure}.

Calcium ferrite (CaFe$_2$O$_4$, recently discovered as the mineral harmunite~\cite{galuskina2014harmunite}) belongs to the orthorhombic crystal system (space group \emph{Pnma}~\cite{decker1957structure}).  Calcium has eight-fold coordination and ferric iron populates two crystallographically nonequivalent octahedra with different degrees of distortion.  These two different octahedra form infinite double chains along the $c$-axis (in \emph{Pbnm} settings). Each chain contains octahedra of only one type. Within the chains octahedra share edges and different chains are linked by common vertices of octahedra. At ambient pressure calcium ferrite undergoes a phase transition to an antiferromagnetic state bellow 200~K~\cite{corliss1967magnetic}. In previous work~\cite{merlini2010letter} an isosymmetric phase transition was observed above 50 GPa and explained by a spin transition of trivalent iron. Recently, this was associated with a Mott transition~\cite{greenberg2013mott}.

Andradite (Ca$_3$Fe$_2$[SiO$_4$]$_3$) is a member of the garnet family. In this garnet structure, calcium is located in the cubic site (X-site) and ferric iron fully occupies the octahedral position (Y-site). Spin crossover in andradite takes place between 60 GPa and 70 GPa~\cite{friedrich2014pressure}.

Here we present our new high-pressure single-crystal MS data for FeBO$_3$, Fe$_2$O$_3$, and\\ Fe$_3$(Fe$_{1.766(2)}$Si$_{0.234(2)}$)(SiO$_4$)$_3$ and analyze the behavior of hyperfine parameters at spin transitions of ferric iron. We also discuss possible structural changes coupled with spin transitions, analyze the compressibility of the ferric iron octahedron, and the relationship between the isomer shift and the polyhedron volume.

The manuscript is organized as follows. First, new experimental data are presented, followed by a detailed analysis of hyperfine parameter behavior at the spin transition. Then, the compressibility of the ferric iron octahedron is determined and possible structural changes coupled with the spin transition are discussed. Finally, the dependence of the center shift on polyhedron volume is examined.

\section{Experimental description}

Iron borate single crystals were grown from a $^{57}$Fe$_2$O$_3$--B$_2$O$_3$--PbO/PbF$_2$ flux as described in Ref.~\onlinecite{kotrbova1985growth}. Hematite single crystals studied in the present work were taken from the same synthesis batch as the crystals used in Ref.~\onlinecite{bykova2016structural}. Skiagite-iron-majorite single crystals were synthesized in a multi-anvil apparatus at 9.5~GPa and 1100~°C from a powdered mixture of chemically pure oxides: $^{57}$Fe$_{1-x}$O, $^{57}$Fe$_2$O$_3$ and SiO$_2$~\cite{ismailova2015high}.

High-quality single crystals with an average size of $\sim30\times30\times5$~$\mu$m$^3$ for iron borate and hematite, and $\sim10\times10\times5$~$\mu$m$^3$ for skiagite-iron-majorite were pre-selected on a three-circle Bruker diffractometer equipped with a SMART APEX CCD detector and a high-brilliance Rigaku rotating anode (Rotor Flex FR-D, Mo-K$_\alpha$ radiation) with Osmic focusing X-ray optics.

For pressure generation BX90 DACs~\cite{kantor2012bx90} were used. The size of diamond culets was 250 $\mu$m. Rhenium gaskets were pre-indented down to a thickness of $\approx30\:\mu$m and holes of $\approx130\:\mu$m diameter were drilled in the center of indents. The cells with sample and a small ruby sphere placed in the pressure chamber  were loaded with Ne (Ar for FeBO$_3$ single-crystal experiment) by means of a gas-loading system~\cite{kurnosov2008novel}. Pressure in the pressure chamber was estimated before and after MS measurements by the ruby fluorescence method~\cite{dewaele2008compression} and an average value was taken and the deviation was included in the uncertainty. For heating of Fe$_2$O$_3$ we used a portable double-sided laser-heating system with infrared laser~\cite{kupenko2012portable} ($\lambda=1071$~nm) in continuous mode. In this case KCl was used as a pressure transmitting and thermal insulation medium.

M\"{o}ssbauer absorption spectra of powdered FeBO$_3$ were collected using a conventional WissEL spectrometer in constant-acceleration mode with a nominal 10~mCi $^{57}$Co(Rh) point source at 19~°C. The folded spectra consist of 256 channels. The single-crystal M\"{o}ssbauer experiments with iron borate, hematite and skiagite were performed at ambient temperature (22~°C) at the Nuclear Resonance beamline (ID18)~\cite{ruffer1996nuclear} at the European Synchrotron Radiation Facility (ESRF) using the synchrotron M\"{o}ssbauer source (SMS)~\cite{potapkin201257fe}. The SMS is the pure nuclear reflection of a $^{57}$FeBO$_3$ crystal which is mounted on a velocity transducer and operated in sinusoidal mode. The SMS is linearly polarized due to polarization of the synchrotron radiation. In this case the folded spectra contain 512 channels. The average beam spot size for these experiments was $8\times13\:\mu$m$^2$. The line width of the SMS was determined before and after collection of each spectrum of the samples by measurements of the reference single line absorber (K$_2$Mg$^{57}$Fe(CN)$_6$). All experiments were conducted in transmission geometry.

The M\"{o}ssbauer spectra were fitted using \emph{MossA} software~\cite{prescher2012mossa}, version 1.01a. Spectra from the conventional radioactive source were processed using pseudo-Voigt line shapes. For processing of spectra collected with SMS, a transmission integral fit was used assuming a Lorentzian-squared line shape of the SMS and a Lorentzian line shape of the absorber. All center shifts were calibrated relative to $\alpha$-Fe at ambient conditions.

The visualization of the structures and calculation of the distortions of  polyhedra were performed using VESTA software~\cite{momma2011vesta}, version 3.3.2. Calculations of the equations of state (EoS) were performed using EosFit7 \cite{angel2014eosfit7c}.

\section{Results}
\subsection{\label{Iron_borate_experiment}Iron borate}

\begin{figure*}[h]
\includegraphics[width=17.8cm, keepaspectratio=true]{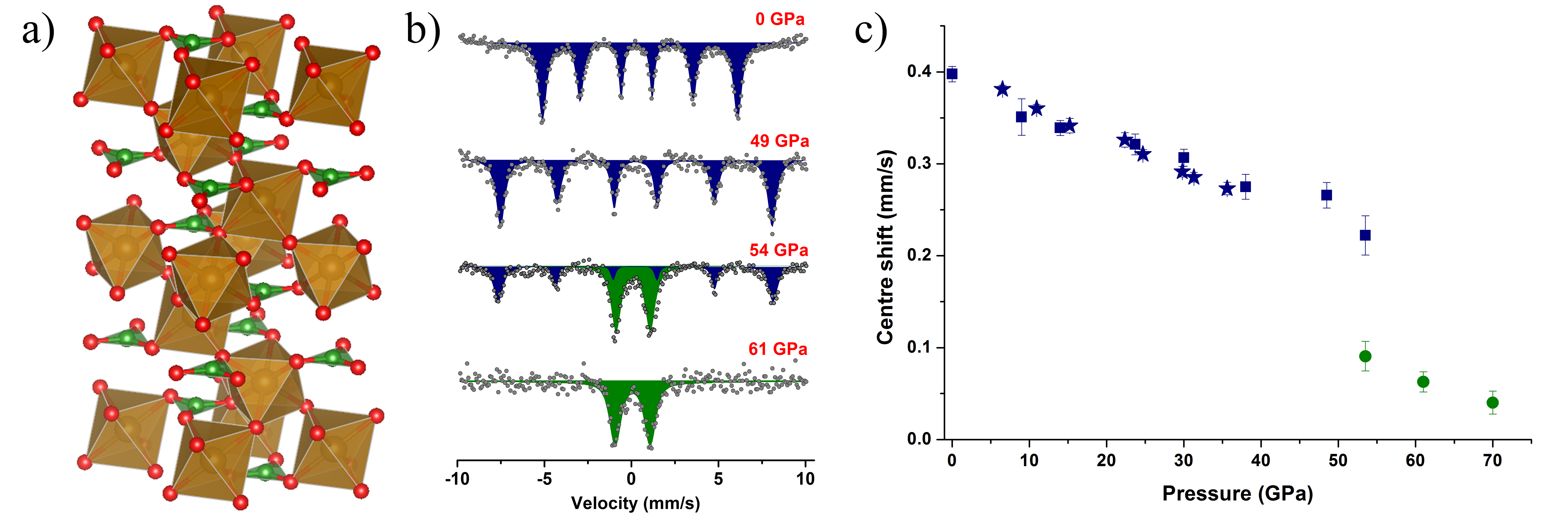}
\caption{\label{Iron_borate_structure}a) Structure of FeBO$_3$ comprised by corner-shared Fe$^{3+}$O$_6$ octahedra and BO$_3$ triangles perpendicular to the 3-fold axis. b) Pressure evolution of iron borate M\"{o}ssbauer spectra. At ambient pressure the spectrum is a single magnetic sextet (blue) that demonstrates increasing $H_{hf}$ at compression up to 54~GPa, then a paramagnetic doublet (green) appears that can be attributed to LS Fe$^{3+}$. At 61~GPa only the paramagnetic component remains. c) Doublet has center shift lower by $\approx0.13$~mm/s than sextet. The squares and circles correspond to the experiment with powder sample measured using a radioactive M\"{o}ssbauer source and the stars are data from the single-crystal experiment measured using SMS. Hereafter, if error bars are not visible they are smaller than the symbol size.}
\end{figure*}

Recent high-pressure single-crystal XRD experiments with FeBO$_3$~\cite{Greenberg_be_published} showed a sharp transition with large volume discontinuity at $50\pm2$~GPa\footnote{hereafter the error notation $A\pm\sigma$ means that the probability density of the error has continuous uniform distribution over the interval $[A-\sigma,A+\sigma]$, as distinct from the notation $A(\sigma)$ where the Gaussian distribution of the error is implied and $\sigma$ is one standard deviation.} and confirmed earlier supposition~\cite{gavriliuk2002equation,parlinski2002structural} that it is isosymmetric phase transition. The single crystal XRD data provide important information about geometrical characteristics of the Fe$^{3+}$O$_6$ octahedron. At ambient conditions the iron octahedron is slightly trigonally elongated along the 3-fold axis. Under compression this distortion diminishes and completely disappears around 36~GPa.   At the transition, the octahedron volume decreases by 10.5~\%, which is a clear indication of the transition to LS state. The octahedron remains ideal across the transition and it is only above 56~GPa that a trigonal compression along the 3-fold axis begins to grow slowly~\cite{Greenberg_be_published}.

\begin{figure*}[t]
\includegraphics[width=17.8cm, keepaspectratio=true]{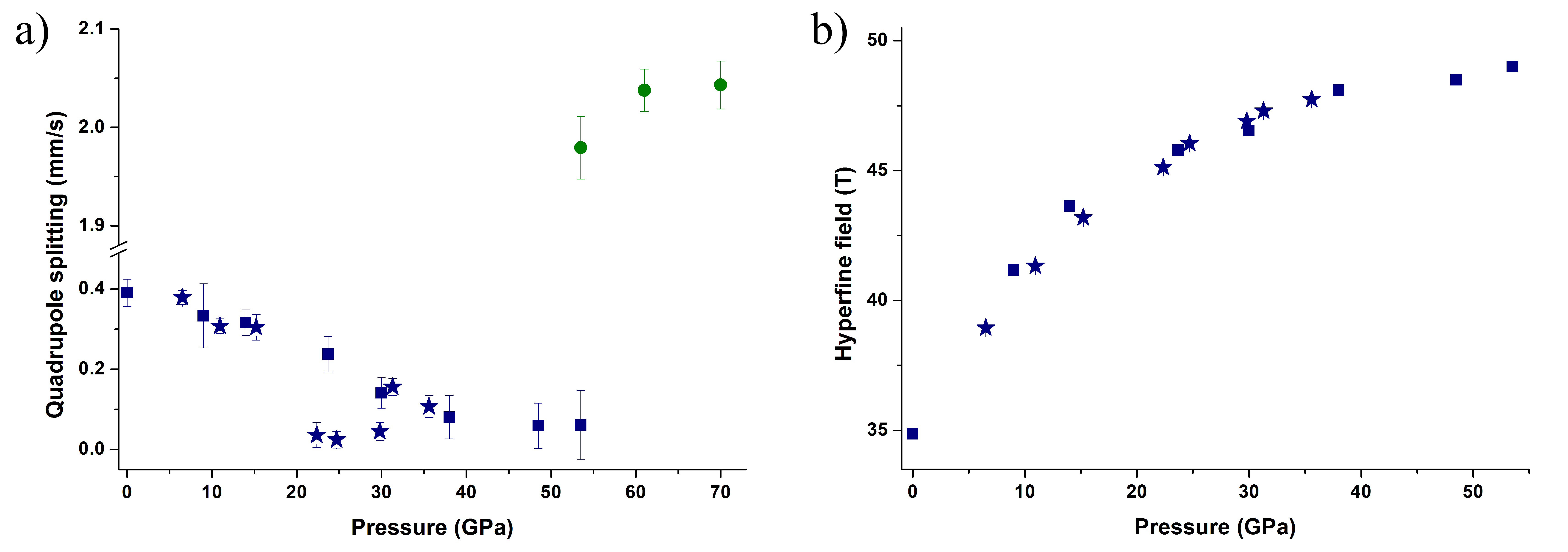}
\caption{\label{FeBO3 moessbauer}a) Effect of pressure on modulus of FeBO$_3$ quadrupole splitting. The values for the doublet (green) and sextet (blue) differ by one order of magnitude at 54~GPa. b)~The growth of hyperfine field over the entire pressure range indicates an increase of N\'{e}el temperature of FeBO$_3$. The squares and circles correspond to the experiment with powder sample measured using a radioactive M\"{o}ssbauer source and the stars are data from the single-crystal experiment measured using SMS. The colors conform to those in Figs.~\ref{Iron_borate_structure}b and~\ref{Iron_borate_structure}c.}
\end{figure*}

M\"{o}ssbauer spectra of iron borate were collected from ambient pressure to 70~GPa. The evolution of the M\"{o}ssbauer spectra is shown in Fig.~\ref{Iron_borate_structure}b. At ambient conditions the spectrum consists of one magnetic component with the following hyperfine parameters: center shift~(CS) $\delta_{CS}=0.398(8)$~mm/s, quadrupole shift $\varepsilon=0.098(9)$~mm/s\footnote{in the literature, $2\varepsilon$ is often referred to as quadrupole shift} and hyperfine magnetic field $H_{hf}=34.86(6)$~T. Up to 48.5~GPa there are no substantial changes of the spectrum, only an increase of hyperfine field (Fig.~\ref{FeBO3 moessbauer}b) and decrease of CS (Fig.~\ref{Iron_borate_structure}c).

The quadrupole shift gradually decreases with increasing pressure and becomes close to zero above 30~GPa. Such behavior can be understood if one takes into account the above-mentioned changes in the octahedron distortion. With pressure increase the trigonal distortion decreases and becomes zero above 35~GPa. This means that there is no lattice contribution to the electric field gradient (EFG) from the first coordination sphere. However, there is still a lattice contribution from next coordination spheres as the iron Wyckoff position does not have cubic symmetry. The quadrupole shift alters accordingly. Indeed, $\varepsilon$ above 35~GPa is small but non-zero.

The iron ions are on the 3-fold axis; therefore the EFG should be symmetric, i.e., the asymmetry parameter $\eta=0$ and the principal component $V_{zz}$ of the EFG tensor is collinear with the 3-fold axis~\cite{cohen1957quadrupole}. As the magnetic moments lie in the plane perpendicular to the 3-fold axis and remain in the plane until the spin transition occurs~\cite{troyan2001transition,sarkisyan2002magnetic}, the angle $\theta$ between $H_{hf}$ and $V_{zz}$ of the EFG remains constant, and we can convert $\varepsilon$ to quadrupole splitting~(QS)~$\Delta$ to have values for HS and LS states that can be compared. In our case the quadrupole shift in the high-field condition ($g_N\mu_NH_{hf}\gg eQV_{zz}/2$) is~\cite{guetlich2010moessbauer}
\begin{equation}\label{quadrupole_shift}
|\varepsilon|=\frac{eQV_{zz}}{2}\left(\frac{3\cos^2\theta-1}{4}\right)=\Delta\left(\frac{3\cos^2\theta-1}{4}\right),
\end{equation}
where $Q$ is the quadrupole moment of the nucleus and $e$ is the elementary charge. For $\theta=90^\circ$, $|\varepsilon|=\Delta/4$ and the projection of the EFG along the field is $V_{xx}$, which has an opposite sign compared to $V_{zz}$ (as the EFG tensor is traceless). Therefore, as the quadrupole shift is positive, the $V_{zz}$ is negative. The respective $|\Delta|$ is plotted in Fig.~\ref{FeBO3 moessbauer}a.

At 54(1)~GPa a new nonmagnetic doublet appears with $\delta_{CS}=0.09(2)$~mm/s ($\approx0.13$~mm/s lower than the CS of the magnetic component, Fig.~\ref{Iron_borate_structure}c) and $|\Delta|=1.98(3)$~mm/s, which is one order of magnitude larger than $\Delta$ before the transition (Fig.~\ref{FeBO3 moessbauer}a). Parameters of the doublet correspond to the low-spin state of the iron ion, $^2T_{2g}$. At 61~GPa the magnetic sextet disappears completely, and only the doublet is observed up to 70~GPa. Our results are in general agreement with the previous study~\cite{sarkisyan2002magnetic}, although our data show smoother pressure dependencies of the hyperfine parameters, and higher pressure of the transition compared to Ref.~\onlinecite{sarkisyan2002magnetic}. %These differences can be explained by use of less hydrostatic pressure transmitting medium (silicon oil \emph{vs} Ne in our experiments)  and a too large crystal (80~$\mu$m in diameter when the hole size was 100~$\mu$m) in~\cite{sarkisyan2002magnetic} that might result in substantial pressure difference across the crystal and its bridging.

\subsection{\label{hematite_results}Iron sesquioxide}

\begin{figure*}[h]
\includegraphics[width=17.8cm, keepaspectratio=true]{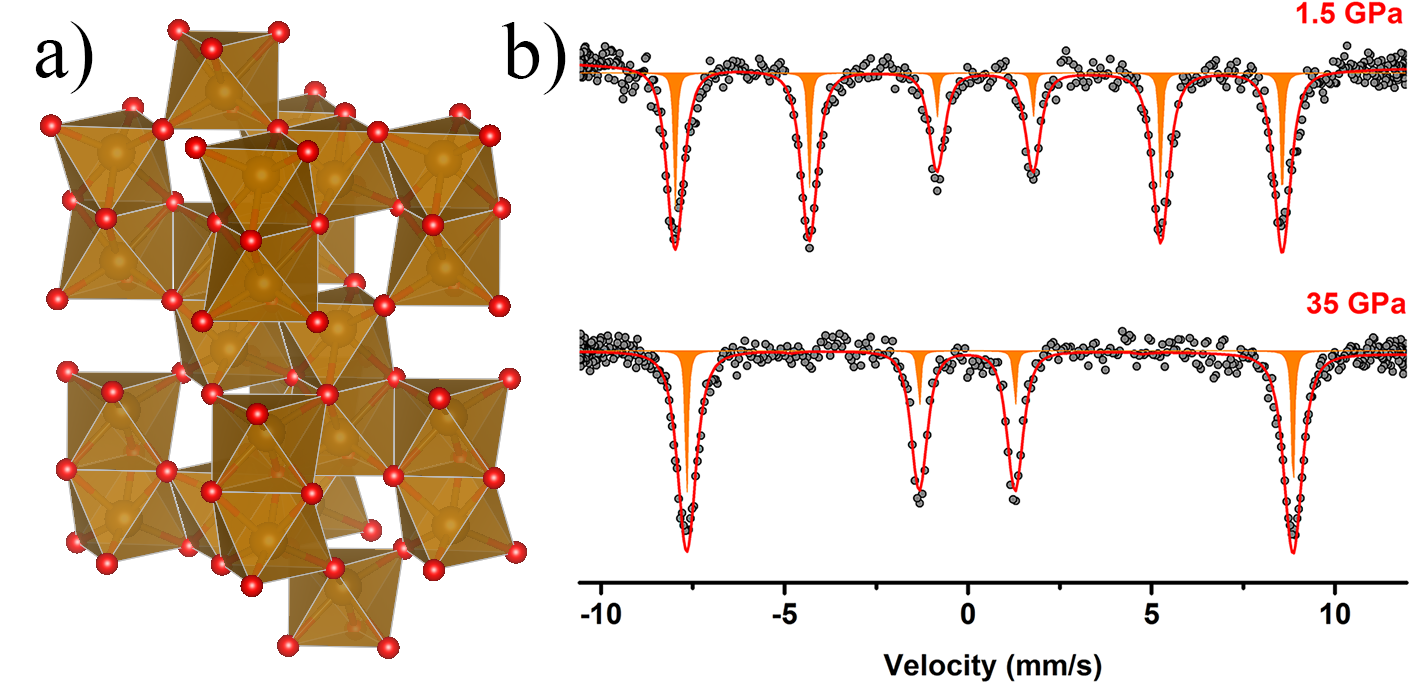}
\caption{\label{Morin_transition}a)~Structure of Fe$_2$O$_3$ (hematite) at ambient conditions consisting of Fe$^{3+}$O$_6$ octahedra. b)~M\"{o}ssbauer spectrum of hematite showing a single magnetic sextet of HS Fe$^{3+}$. The spin-flop Morin transition results in a disappearance of the 2nd and 5th lines due to the particular orientation of the single crystal relative to the incident beam.}
\end{figure*}

\begin{figure*}[h]
\includegraphics[width=17.8cm, keepaspectratio=true]{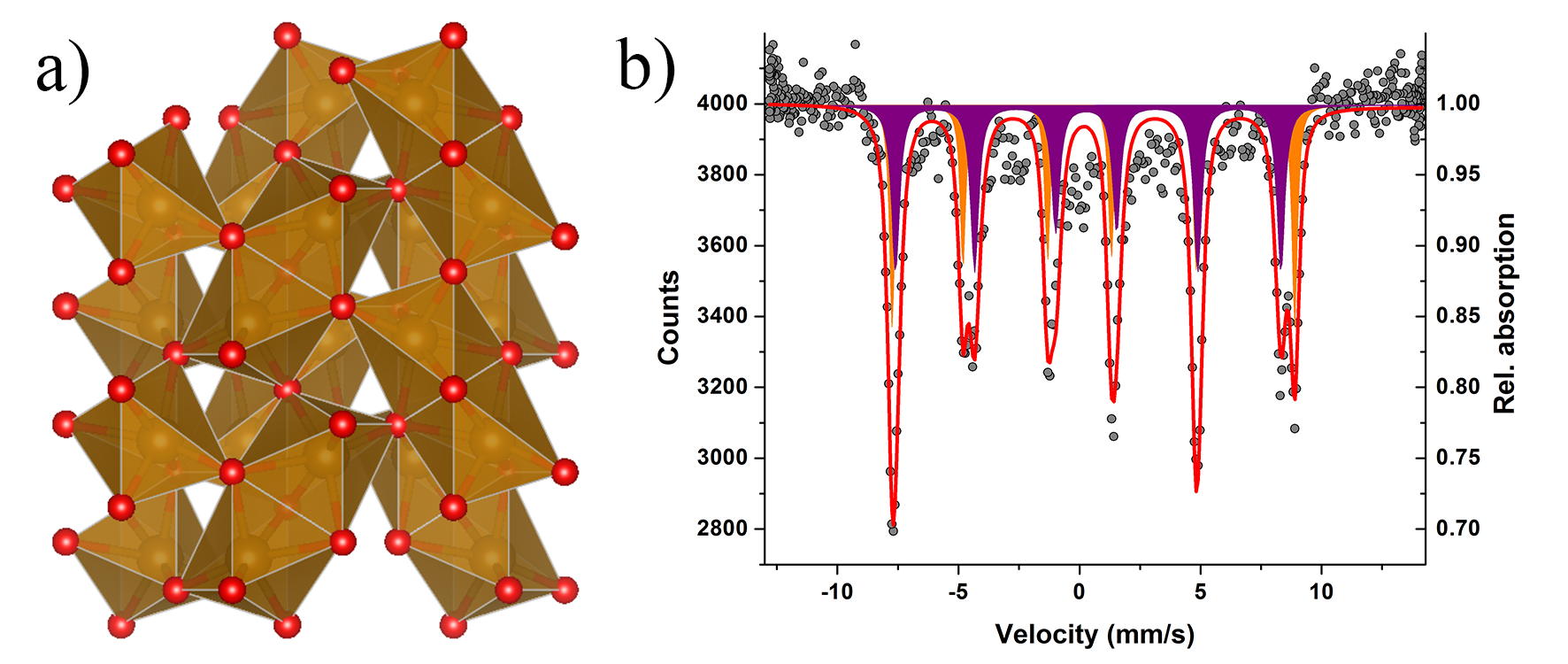}
\caption{\label{iota-Fe2O3}a)~Crystal structure of $\iota$-Fe$_2$O$_3$ with iron octahedra as in hematite but with a different packing motif. b)~M\"{o}ssbauer spectrum after laser heating of $\alpha$-Fe$_2$O$_3$ ($\sim1400$~K) and quenching at 29(3)~GPa showing superposition of two magnetic sextets. The orange and purple components are $\alpha$-Fe$_2$O$_3$ and $\iota$-Fe$_2$O$_3$, respectively.}
\end{figure*}

\begin{figure*}[t]
\includegraphics[width=17.8cm, keepaspectratio=true]{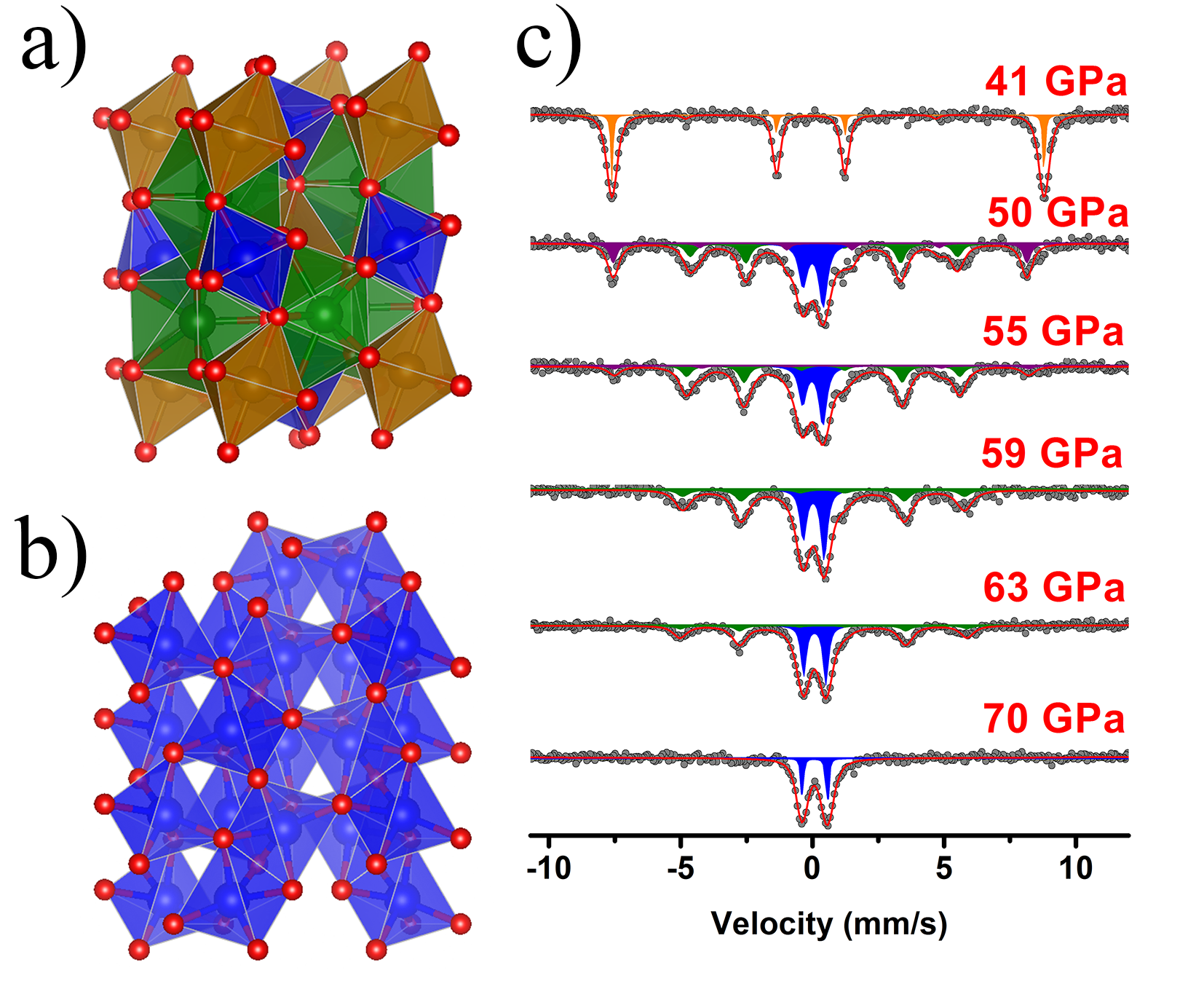}
\caption{\scriptsize\label{Hematite structure and evolution}Structures of high-pressure Fe$_2$O$_3$ polymorphs~\cite{bykova2013novel,bykova2016structural} and evolution of M\"{o}ssbauer spectra at the respective transitions. a)~At 54~GPa $\alpha$-Fe$_2$O$_3$ undergoes a phase transition to $\zeta$-Fe$_2$O$_3$ with distorted perovskite-type structure that consists of two types of FeO$_6$ octahedra with different volumes (B-position, small octahedron is blue) and FeO$_8$ bicapped trigonal prisms (A-position). b)~At 67~GPa $\zeta$-Fe$_2$O$_3$ transforms to $\theta$-Fe$_2$O$_3$ built of FeO$_6$ trigonal prisms. c)~ At 50~GPa the M\"{o}ssbauer spectrum changes to a superposition of the magnetic sextet of $\iota$-Fe$_2$O$_3$ (purple component) and the signal of $\zeta$-Fe$_2$O$_3$: doublet (blue) and sextet (green, $H_{hf}=31.4$~T). The $\iota$-Fe$_2$O$_3$ component disappears rapidly with increasing pressure and is absent above 55~GPa. The intensity of the sextet with smaller field starts to decrease at 60~GPa and by 70~GPa the spectrum consists only of one doublet.}
\end{figure*}

\begin{figure*}[t]
\includegraphics[width=17.8cm, keepaspectratio=true]{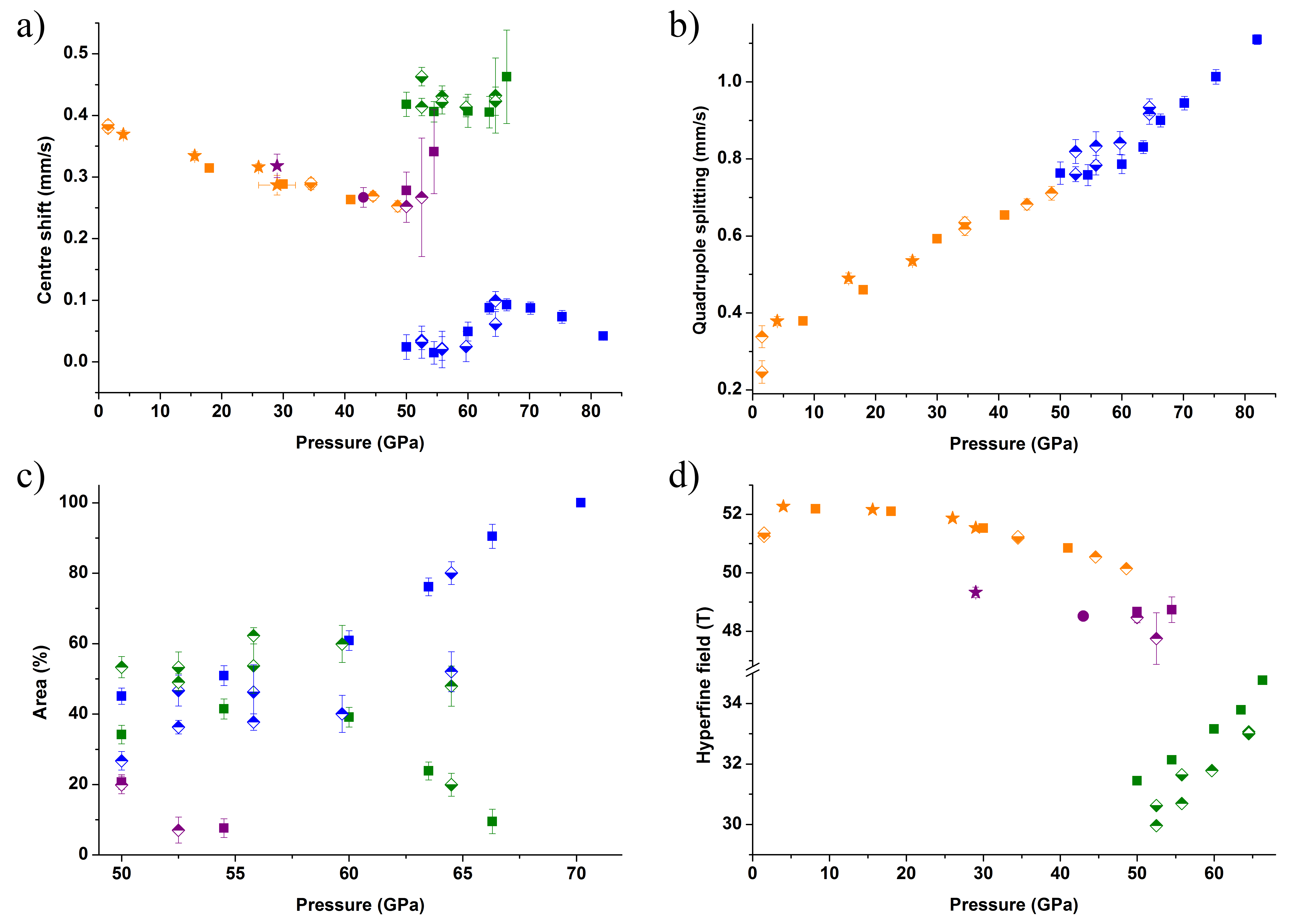}
\caption{\scriptsize{\label{Hematite_hyperfine_parameters}Pressure dependence of Fe$_2$O$_3$ hyperfine parameters. a)~The doublet (blue) of $\zeta$-Fe$_2$O$_3$ has a CS that is lower by $\approx0.2$~mm/s than the value for HS Fe$^{3+}$ before the transition (orange). The CS of the new sextet (green) is larger than the value for hematite at ambient pressure. b)~$\Delta$ of the doublet is approximately equal to QS in $\alpha$-Fe$_2$O$_3$ before the transition. c)~The relative amount of components after the transition to $\zeta$-phase shows that $\iota$-Fe$_2$O$_3$ (purple) vanishes at 55~GPa, and above 60~GPa the green sextet starts to disappear. At 70~GPa the doublet remains the single component in the spectrum. d)~The hyperfine field of $\iota$-Fe$_2$O$_3$ is lower by 1.5--2 T than values for $\alpha$-Fe$_2$O$_3$ over the entire pressure range. $H_{hf}$ of the green sextet is around 30~T at 50~GPa and increases with increasing pressure. The colors conform to those in Fig.~\ref{Hematite structure and evolution}c. Squares, stars and half-filled diamonds correspond to data from three different DACs. Half-up and half-down filled symbols distinguish two different crystals in the third DAC. The circles at 43~GPa are parameters of $\iota$-Fe$_2$O$_3$ from Ref.~\onlinecite{Kupenko_be_published}.}}
\end{figure*}

M\"{o}ssbauer spectra of Fe$_2$O$_3$ were collected up to 82~GPa. We conducted independent experiments with three DACs (one of them contained two crystals measured separately). Selected spectra are presented in Figs.~\ref{Morin_transition}b,~\ref{iota-Fe2O3}b and~\ref{Hematite structure and evolution}c. The behavior of the hyperfine parameters with pressure is shown in Fig.~\ref{Hematite_hyperfine_parameters}.

The M\"{o}ssbauer spectrum of $\alpha$-Fe$_2$O$_3$ (hematite) is a single magnetic sextet of HS ferric iron (Fig.~\ref{Morin_transition}b). At ambient temperature hematite undergoes a pressure-induced spin-flop Morin transition at $1.7$~GPa~\cite{klotz2013pressure}. At this transition the spins flip~$90^\circ$ from the basal plane to the direction collinear with the 3-fold axis without change in the atomic arrangement. The noticeable effect of this transition is the disappearance of the 2nd and 5th lines of the sextet in the M\"{o}ssbauer spectrum (Fig.~\ref{Morin_transition}b) due to the use of a single-crystal sample:  in this particular experiment  the angle between the wave vector of the $\gamma$-ray and the magnetic moments is close to zero after the transition.  The hyperfine magnetic field is increased by~1.0(1)~T~(Fig.~\ref{Hematite_hyperfine_parameters}d) and $\varepsilon$ changes from $-0.086(7)$ to $+0.189(6)$~mm/s, which is in good agreement with data for the Morin transition at ambient pressure~\cite{tobler1981investigation}.

There is no significant change of hematite component with further compression up to 49~GPa. The $\delta_{CS}$ expectedly decreases~(Fig.~\ref{Hematite_hyperfine_parameters}a) and $H_{hf}$ reduces from 52.26(5)~T to 50.15(6)~T~(Fig.~\ref{Hematite_hyperfine_parameters}d). Note that the latter behavior is not related to the reduction of the N{\'e}el temperature~\cite{Kupenko_be_published}, but is caused by a decrease of the $H_{hf}$ saturation value  with pressure increase (a similar behavior was observed in FeBO$_3$~\cite{gavriliuk2005high}). Although single-crystal XRD experiments did not detect any significant changes in the octahedral distortion up to 25~GPa~\cite{schouwink2011high}, the $\varepsilon$ value almost doubles up to 0.355(9)~mm/s at 49~GPa. Since iron is located on the 3-fold axis in $\alpha$-Fe$_2$O$_3$, we can convert the quadrupole shift to $\Delta$ (similar to FeBO$_3$ in section~\ref{Iron_borate_experiment}). The result is plotted in Fig.~\ref{Hematite_hyperfine_parameters}b.

According to data in Ref.~\onlinecite{ito2009determination}, $\alpha$-Fe$_2$O$_3$ is metastable above 40~GPa at room temperature (RT) and orthorhombic $\iota$-Fe$_2$O$_3$ (space group $Pbcn$) becomes the stable phase. This polymorph has a Rh$_2$O$_3$-II structure type with a single crystallographic position of octahedrally-coordinated iron and different from the corundum packing motif. In this phase two iron octahedra form a building block sharing a common face. Connected via common vertices, this subunit forms a layer with herringbone-type packing and, in turn, the layers are interconnected through common edges of iron octahedra (Fig.~\ref{iota-Fe2O3}a).

\emph{Ab initio} calculations for Al$_2$O$_3$~\cite{xu2010first} showed that the kinetic barrier between corundum and Rh$_2$O$_3$-II structures depends on whether the transition is forward (corundum $\rightarrow$ Rh$_2$O$_3$-II) or backward (Rh$_2$O$_3$-II $\rightarrow$ corrundum). The forward barrier is almost pressure independent, while the backward one strongly decreases with decompression, so the phase with Rh$_2$O$_3$-II structure is not recoverable. Our experiments and decompression data in Ref.~\onlinecite{[][; in this study the $\iota$-Fe$_2$O$_3$ signal was erroneously interpreted as a double-perovskite phase in the insulating state according to our interpretation]greenberg2017theoretical} show that these conclusions are also valid for the Fe$_2$O$_3$ system.

In order to determine hyperfine parameters of $\iota$-Fe$_2$O$_3$ we performed a separate experiment (stars in Fig.~\ref{Hematite_hyperfine_parameters}) with laser heating ($\sim1400$~K) of hematite at 29~GPa. The M\"{o}ssbauer spectrum of the quenched sample shows two sextets with similar hyperfine parameters, one belonging to $\alpha$-Fe$_2$O$_3$ (orange in Fig.~\ref{iota-Fe2O3}b) and the second being the signal from $\iota$-Fe$_2$O$_3$ (purple in Fig.~\ref{iota-Fe2O3}b). The hyperfine parameters of the purple sextet are $\delta_{CS}=0.32(2)$~mm/s, $\varepsilon=0.04(2)$~mm/s, and $H_{hf}=49.3(2)$~T. They are in excellent agreement with parameters of pure $\iota$-Fe$_2$O$_3$ at 43~GPa~\cite{Kupenko_be_published} (circle in Fig.~\ref{Hematite_hyperfine_parameters}a~and~d). The transmission integral fit gives a 0.4~mm/s line width for $\iota$-Fe$_2$O$_3$ that is distinct from the natural absorber line width (0.097~mm/s) of hematite. The CS values of $\alpha$-Fe$_2$O$_3$ and $\iota$-Fe$_2$O$_3$ are the same within experimental uncertainty (Fig.~\ref{Hematite_hyperfine_parameters}a) as iron occupies an octahedron with similar volumes in both structures. The main distinctions of the $\iota$-Fe$_2$O$_3$ sextet are practically zero quadrupole shift over the entire pressure range and $H_{hf}$ which is lower by 1.5--2~T compared to hematite values (Fig.~\ref{Hematite_hyperfine_parameters}d).

With further compression of $\alpha$-Fe$_2$O$_3$ the M\"{o}ssbauer spectrum changes drastically at 50(1)~GPa (Fig.~\ref{Hematite structure and evolution}c):  it becomes a superposition of the $\iota$-Fe$_2$O$_3$ component (purple), and a new magnetic sextet ($\delta_{CS}=0.42(2)$~mm/s, $\varepsilon=0.01(2)$~mm/s, $H_{hf}=31.4(2)$~T, green in Fig.~\ref{Hematite structure and evolution}c) and doublet ($\delta_{CS}=0.024(20)$~mm/s and $\Delta=0.76(3)$~mm/s, blue in Fig.~\ref{Hematite structure and evolution}c), while the signal from $\alpha$-Fe$_2$O$_3$ disappears completely. The line width of the new components is 0.4~mm/s.

As single-crystal XRD experiments revealed a phase transition to triclinic $\zeta$-Fe$_2$O$_3$ (space group $P\bar{1}$) at 54(1)~GPa~\cite{bykova2013novel,bykova2016structural}, the new components should be associated with the $\zeta$-phase. The structure of the $\zeta$-Fe$_2$O$_3$ modeled (in monoclinic symmetry~\cite{bykova2013novel}) as a double-perovskite type, consists of bicapped trigonal prisms (A-site) and two kinds of alternating octahedra (B$^{\prime}$-~and B$^{\prime\prime}$-sites). The octahedra have the same abundance but differ from each other in volume (8.6 and 7.5~{\AA}$^3$ at 54~GPa). The discrepancy with XRD data~\cite{bykova2013novel} in transition pressure is apparently related to the use of different methods for pressure determination (ruby fluorescence \emph{vs} position of the (111) XRD line of Ne).

The behavior of hyperfine parameters is in good agreement in all conducted experiments (Fig.~\ref{Hematite_hyperfine_parameters}a,b,d). However, in two independent experiments the observed difference between the relative amount of components in $\zeta$-Fe$_2$O$_3$ after the transition from $\alpha$-Fe$_2$O$_3$ is significantly larger than statistical uncertainty (Fig.~\ref{Hematite_hyperfine_parameters}c). This discrepancy and small difference in $H_{hf}$ values of the green sextet  (Fig.~\ref{Hematite_hyperfine_parameters}d) might indicate that electronic states of iron ions in $\zeta$-Fe$_2$O$_3$ are sensitive to non-hydrostatic stress. Despite this difference, the data show the same general trends.

With further compression $\iota$-Fe$_2$O$_3$ vanishes quickly and is absent in spectra above 55~GPa (Fig.~\ref{Hematite structure and evolution}c and Fig.~\ref{Hematite_hyperfine_parameters}c). There is an approximately constant ratio between areas of the remaining sextet and non-magnetic doublet up to 60~GPa. Above this pressure the green sextet also starts to gradually disappear and the CS and QS pressure dependences of the doublet have inflections around 60~GPa (Fig.~\ref{Hematite_hyperfine_parameters}a,b). The quadrupole shift of the $\zeta$-Fe$_2$O$_3$ magnetic sextet remains zero within statistical uncertainty during compression. Note that the increase of the hyperfine field of the green sextet (Fig.~\ref{Hematite_hyperfine_parameters}d) could signify that $H_{hf}$ is far from its saturation value,  and therefore the magnetic critical point in $\zeta$-Fe$_2$O$_3$ is considerably lower than that of hematite before the transition (the data in Ref.~\onlinecite{greenberg2017theoretical} confirm this conclusion). At 70~GPa the doublet remains the only component of the spectra, in agreement with a single structural position of iron in orthorhombic $\theta$-Fe$_2$O$_3$~\cite{bykova2016structural}. In this phase the line width decreases to 0.13~mm/s. Upon decompression from 61 to 50~GPa, M\"{o}ssbauer spectra reveal a fully reversible behavior ($\iota$-Fe$_2$O$_3$ starts to appear again at 50~GPa).

The interpretation of the M\"{o}ssbauer components of $\zeta$-Fe$_2$O$_3$ within the framework of the proposed double-perovskite structure~\cite{bykova2013novel} is as follows. The peculiar feature of the green sextet is the CS value which is even higher than that of hematite at ambient conditions (Fig.~\ref{Hematite_hyperfine_parameters}a). Since the isomer shift of iron increases with increasing iron coordination number~\cite{menil1985systematic}, we can conclude that this sextet corresponds to HS Fe$^{3+}$ in the bicapped trigonal prisms, i.e., the A-site (with 8 oxygen neighbors). Accordingly, the blue doublet is assigned to the signal from the octahedral B-site and corresponds to the LS Fe$^{3+}$, as this component is non-magnetic and has CS lower by 0.23(2)~mm/s than HS ferric iron before the transition (Fig.~\ref{Hematite_hyperfine_parameters}a).

The interpretation connected with the double-perovskite structure encounters a number of difficulties, however. Firstly, the model structure has two types of octahedra with different volumes (corresponding to HS and LS states of the iron ion). One would therefore expect that $\zeta$-Fe$_2$O$_3$ should give three components in M\"{o}ssbauer spectra. Contrary to that, our experimental data do not show any sign of an additional component. We emphasize that the $\iota$-Fe$_2$O$_3$ sextet (Fig.~\ref{Hematite structure and evolution}c) cannot be this third component due to high $H_{hf}$ (Fig.~\ref{Hematite_hyperfine_parameters}d), because the $\zeta$-Fe$_2$O$_3$ magnetic critical temperature is significantly lower than that in $\alpha$-Fe$_2$O$_3$ and $H_{hf}$ is far from saturation~\cite{greenberg2017theoretical}. Moreover, the volumes of both octahedra in $\zeta$-Fe$_2$O$_3$ at 54~GPa (8.6 and 7.5~{\AA}$^3$) are too small compared to the rest of the examined data (see section~\ref{Oct_volume_section}).

Secondly, according to the model structure of the $\zeta$-Fe$_2$O$_3$ phase~\cite{bykova2013novel,bykova2016structural}, one could expect that after the HS$\rightarrow$LS transition in the octahedra, the M\"{o}ssbauer spectrum will have two components with area ratio~1:1 (HS iron in bicapped trigonal prisms and LS iron in octahedra). The subsequent transition to $\theta$-Fe$_2$O$_3$ should result in an abrupt vanishing of one component. However, our data clearly show that the green sextet disappears gradually (Fig.~\ref{Hematite structure and evolution}c), and at 65~GPa the area ratio between the doublet and the sextet is around 4:1 (Fig.~\ref{Hematite_hyperfine_parameters}c).

Thirdly, it is not clear why a strong change of the coordination polyhedron (octahedron to trigonal prism) at the $\zeta$-Fe$_2$O$_3$ to $\theta$-Fe$_2$O$_3$ transition  does not lead to discontinuities in the hyperfine parameters (Fig.~\ref{Hematite_hyperfine_parameters}a,b). Moreover, the data show that QS values of $\alpha$-Fe$_2$O$_3$ and $\zeta$-Fe$_2$O$_3$ doublets are almost the same at 50~GPa (Fig.~\ref{Hematite_hyperfine_parameters}b). Such behavior is at variance with the general trend of the increase of $\Delta$ at the HS$\longrightarrow$LS transition of Fe$^{3+}$ in the octahedra~\cite{nihei2007spin}.

To overcome these disagreements, we note that $\iota$-Fe$_2$O$_3$ and $\theta$-Fe$_2$O$_3$ have the same packing motif (Fig.~\ref{iota-Fe2O3}a and~\ref{Hematite structure and evolution}b), the only difference being the type of iron polyhedron (octahedron \emph{vs} trigonal prism). Moreover, at the $\alpha$- to $\zeta$-Fe$_2$O$_3$ transition our experiments show a simultaneous appearance of $\iota$-Fe$_2$O$_3$ together with the $\zeta$-phase (Fig.~\ref{Hematite structure and evolution}c and~\ref{Hematite_hyperfine_parameters}c). Data in Ref.~\onlinecite{greenberg2017theoretical} clearly demonstrate that $\zeta$-Fe$_2$O$_3$ transforms to $\iota$-Fe$_2$O$_3$ under decompression, not to $\alpha$-Fe$_2$O$_3$. These facts indicate that $\zeta$-Fe$_2$O$_3$ is an intermediate phase in the reconstructive transition between $\iota$-Fe$_2$O$_3$ and $\theta$-Fe$_2$O$_3$. The stabilization of the intermediate phase over a finite pressure range and the complex dynamics of this reconstructive transition arise due to changes of iron spin state (and, accordingly, ionic radius) across it.

The double-perovskite structure (Fig.~\ref{Hematite structure and evolution}a) therefore does not appear to be a suitable candidate for this intermediate phase. Based on our M\"{o}ssbauer data, it is reasonable to assume that the trigonal prisms are already present in the $\zeta$-Fe$_2$O$_3$ structure and associate the blue doublet (Fig.~\ref{Hematite structure and evolution}c) with them. This explains the smooth behavior of the doublet hyperfine parameters  at the transition from $\zeta$- to $\theta$-Fe$_2$O$_3$ (Fig.~\ref{Hematite_hyperfine_parameters}a,b). The green sextet corresponds to HS Fe$^{3+}$ (we cannot say anything about its polyhedron besides that a coordination number higher than six is plausible) and its progressive disappearance above 60~GPa might indicate that changes in the shape of polyhedra and interatomic distances proceed gradually. A theoretical investigation of the possible pathways between $\iota$-Fe$_2$O$_3$ and $\theta$-Fe$_2$O$_3$ structures should clarify this problem.

In concluding this section, it is important to note that trigonal prismatic coordination ($D_{3h}$ symmetry group) does not lead to the habitual splitting of $3d$-levels on $e_g$,~$t_{2g}$ manifolds as octahedral, tetrahedral ($e$,~$t_2$) and cubic environments. Instead, the $d$-levels split into a low-lying $d_{z^2}$ singlet and $d_{x^2-y^2}$,$d_{xy}$ doublet, and high-energy $d_{xz}$,$d_{yz}$ doublet~\cite{huisman1971trigonal}. It is evident that in the trigonal prism, the crystal field does not quench orbital moments at all and it is convenient to use the designations $d_0$, $d_{\pm2}$ and $d_{\pm1}$ for $3d$-orbitals.

In the case of the trigonal prism, the intermediate spin state can be readily stabilized in addition to HS and LS states, as the energies of $d_0$ and $d_{\pm2}$ levels are sensitive to the geometry of the prism~\cite{huisman1971trigonal}. To~exclude this possibility we use the fact that the metal-ligand distance is approximately the same for both octahedral and trigonal prismatic coordination~\cite{huisman1971trigonal}. Comparison of the average Fe-O distance in $\theta$-Fe$_2$O$_3$ (1.82(4)~{\AA} at 74~GPa~\cite{bykova2016structural}) with typical values for LS Fe$^{3+}$ in oxygen octahedra~\cite{bykova2013novel} shows unambiguously that iron ions are in the low-spin state in this phase, in agreement with conclusions based on M\"{o}ssbauer data.

For LS Fe$^{3+}$ in the trigonal prism, there are two possible electronic ground states: non-degenerate $^2A^\prime_1$ (with one electron on the $d_0$ orbital) and the doubly degenerate $^2E^\prime$ state (with three electrons on the $d_{\pm2}$ orbitals). The highly distorted trigonal prisms in $\theta$-Fe$_2$O$_3$~\cite{bykova2016structural} indicate the Jahn-Teller nature of the iron ion and, therefore, support the $^2E^\prime$~electronic state. The peculiarity of $^2E^\prime$ is a huge orbital moment $2\mu_B$ (for a purely ionic state), exceeding the spin contribution to the total magnetic moment. Hence, LS Fe$^{3+}$ in a trigonal prism is substantially different from the octahedral case and this should be taken into account in the analysis of physical properties of $\zeta$- and $\theta$-Fe$_2$O$_3$.

\subsection{Skiagite-iron-majorite solid solution}

\begin{figure*}[h]
\includegraphics[width=17.8cm, keepaspectratio=true]{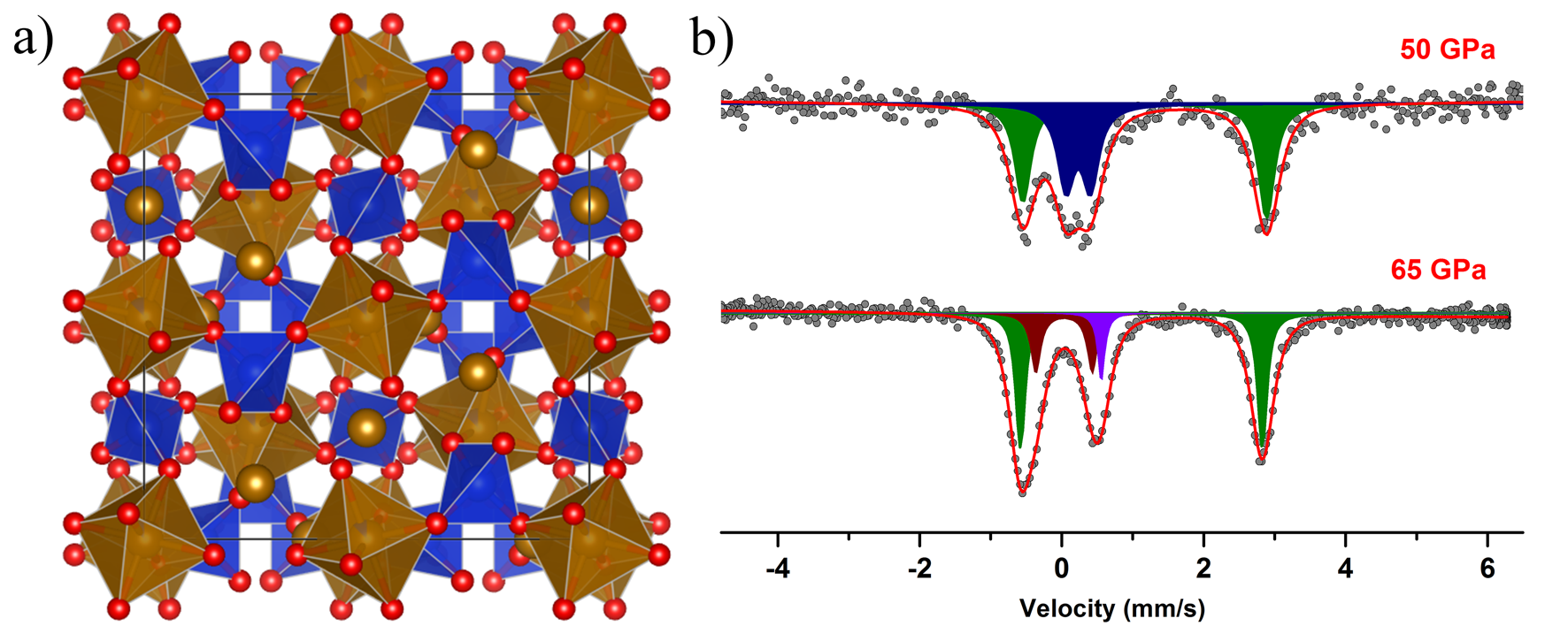}
\caption{\label{Sk_Mj_structure}a)~Skiagite-iron-majorite solid solution with typical garnet structure. Its framework is formed by corner-shared SiO$_4$ tetrahedra (Z-site, blue) and (Fe,Si)O$_6$ octahedra (Y-site, gold), and the distorted cubic voids (X-site) are populated by Fe$^{2+}$. In the pressure range 50~–~60~GPa, skiagite-iron-majorite solid solution undergoes an isosymmetric phase transition with $\approx10$~\% discontinuity of (Fe,Si)O$_6$ volume~\cite{ismailova2017effect}. b) M\"{o}ssbauer spectra of skiagite-majorite solid solution before (50~GPa) and after (65~GPa) spin crossover. The green doublet corresponds to the X-site with HS Fe$^{2+}$, the blue and brown doublets correspond to the Y-site with HS and LS Fe$^{3+}$, respectively, and the violet singlet corresponds to the LS Y-site Fe$^{2+}$.}
\end{figure*}
%The XRD study of Fe$_3$(Fe$_{1.766(2)}$Si$_{0.234(2)}$)(SiO$_4$)$_3$~\cite{ismailova2017effect} show that in the pressure range of 52--62 GPa the volume of the octahedron drops by 9.5~\%, what suggests the spin crossover of iron in this site. To confirm it we performed two independent M\"{o}ssbauer experiments with DACs up to 82~GPa.

Below 50~GPa the M\"{o}ssbauer spectrum of the studied solid solution is a superposition of two paramagnetic doublets corresponding to ferrous iron in the X-site and ferric iron in the Y-site (Fig.~\ref{Sk_Mj_structure}b).  At 1.4~GPa the hyperfine parameters are $\delta_{CS}=1.292(9)$~mm/s, $\Delta=3.49(2)$~mm/s for X-site Fe$^{2+}$ and $\delta_{CS}=0.36(2)$~mm/s, $\Delta=0.25(4)$~mm/s for Y-site Fe$^{3+}$, which is in excellent agreement with~\cite{ismailova2015high}. As the garnet structure is cubic, even in the case of single-crystal M\"{o}ssbauer experiments the doublet components should have equal areas for any orientation~\cite{Spiering1984cubic_crystals}. A small asymmetry of the X-site doublet areas appears due to the Gol'danskii-Karyagin effect (GKE)~\cite{geiger1992combined}. Note that the GKE is not possible for the Y-site, as all diagonal elements of the mean-square displacement tensor are equal due to the symmetry of this position.

The signal from Y-site Fe$^{2+}$ is not detected in the M\"{o}ssbauer spectra, despite the sample containing 23~mol.~\% of iron-majorite. Apparently, the small amount of octahedral divalent iron is not distinguishable within the statistics of data collection. The interpretation that the Y-site doublet (blue, Fig.~\ref{Sk_Mj_structure}b) is the signal from iron in a mixed valence state is implausible. Mixed valence components are observed in compounds where iron polyhedra (participating in intervalence charge transfer) are connected through common edges or faces~\cite{amthauer1984mixed}, which is not the case for iron populating the octahedral position in the garnet structure (see Fig.~\ref{Sk_Mj_structure}a).

The M\"{o}ssbauer spectra do not change substantially with compression up to 50~GPa. Above this pressure the Fe$^{3+}$ component starts to transform into an asymmetric doublet with higher quadrupole splitting (Fig.~\ref{Sk_Mj_structure}b). The changes develop up to 65~GPa, so the transformation pressure range is consistent with that determined from the XRD data~\cite{ismailova2017effect} within experimental error. Between these pressures the Y-site component shows dynamical broadening which indicates interconversion between spin states comparable with the mean lifetime ($\sim10^{-7}$~s) of the excited state of $^{57}$Fe.

\begin{figure*}[h]
\includegraphics[width=17.8cm, keepaspectratio=true]{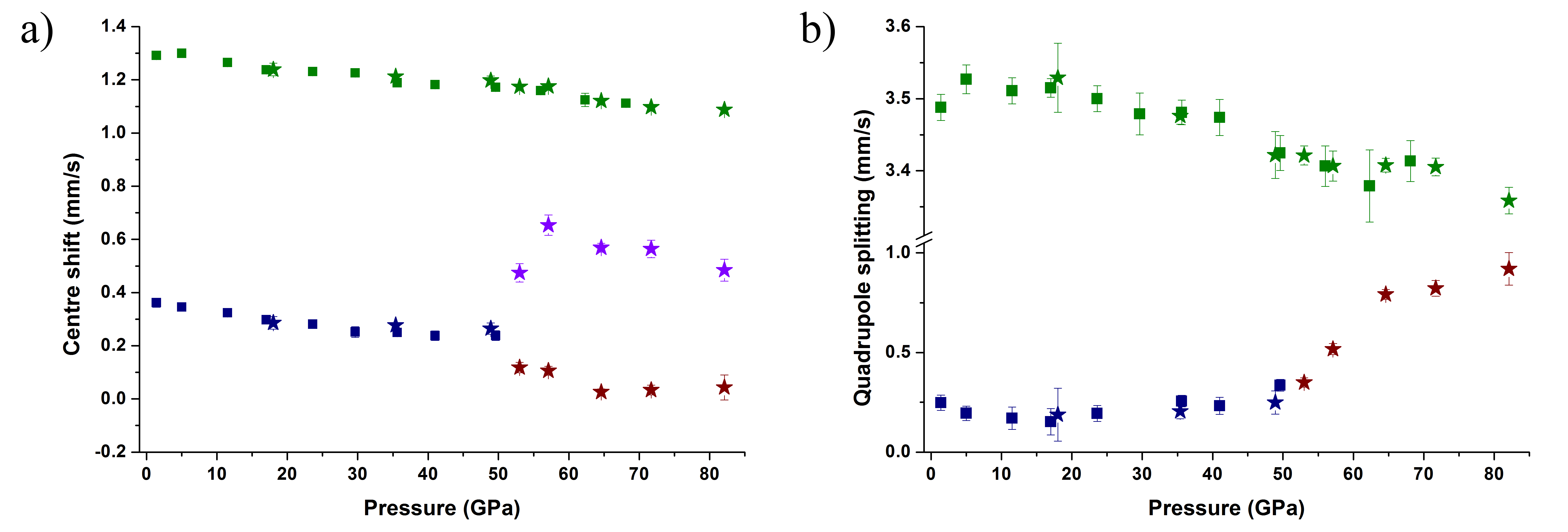}
\caption{\label{Skiagite_moessbauer}Pressure dependence of center shift (a) and quadrupole splitting (b). The hyperfine parameters of X-site Fe$^{2+}$ (green) show monotone behavior over the entire all pressure range. Above 50~GPa, the CS of Y-site Fe$^{3+}$ decreases with simultaneous growth of QS (blue and brown points). The colors conform to those in Fig.~\ref{Sk_Mj_structure}b and the squares and stars correspond to two independent experiments.}
\end{figure*}

Since the structure of skiagite-iron-majorite preserves the cubic symmetry over the entire investigated pressure range, the intensities of the Y-site doublet lines should remain equal after spin crossover~\cite{Spiering1984cubic_crystals}. To describe the new feature (strong asymmetry of the new doublet) one could add a singlet to the more intense line of the new doublet and interpret it as iron in a different electronic state in the octahedral position (Fig.~\ref{Sk_Mj_structure}b, spectrum at 65~GPa). Indeed the value of the center shift of the singlet (0.57(2)~mm/s at 65~GPa) is close to values characteristic for  LS Fe$^{2+}$ in oxygen octahedra \cite{kantor2006spin,cerantola2015high}. This observation could be an indication that octahedral ferrous iron undergoes a HS$\rightarrow$LS transition. Due to collapse of the doublet into a singlet, the dip of this component increases by a factor of two, hence the signal from this ion could become distinguishable.

The asymmetry of the Y-site doublet could also be caused by a dynamical Jahn-Teller effect of LS Fe$^{3+}$, because in the low-spin state, ferric iron has the orbitally degenerate state $^2$T$_{2g}$. There are several equivalent distortions that remove degeneracy, and the iron octahedron can resonate between them. It could lead to non-trivial modifications of the line shape, if the hopping characteristic time between equivalent distorted configurations is close to the characteristic measurement time~\cite{bersuker1968hyp}.

We note, however, that while the proposed explanations of the asymmetry of the doublet described above are ambiguous and further work (for example, low-temperature MS) may be required to fully explain the observations, this uncertainty does not affect our major conclusions in any way. The brown doublet (Fig.~\ref{Sk_Mj_structure}b) remains the signal of trivalent iron in the octahedral site. The hyperfine parameters of ferric iron during the transition change in the following way: at 65~GPa CS drops from 0.24 (at 50~GPa) to 0.03~mm/s and the QS value increases from 0.34 to 0.8~mm/s~(Figs.~\ref{Skiagite_moessbauer}a and~\ref{Skiagite_moessbauer}b). These changes are consistent with typical trends for spin transitions of ferric iron. Hyperfine parameters of divalent iron in the cubic position vary monotonically up to 90~GPa~(Figs.~\ref{Skiagite_moessbauer}a and~b).

\section{Discussion}
\subsection{Hyperfine parameters at spin transitions in Fe$^{3+}$O$_6$ octahedra}

\begin{table*}[h]
\caption{\label{MS_parameters}Hyperfine parameters of ferric iron in the studied compounds before and after the spin transition. All parameters refer to ambient temperature if not given explicitly.}
\begin{ruledtabular}
\begin{tabular}{cccccc}
Compound & $\delta_{HS}-\delta_{LS}$, mm/s & \multicolumn{2}{c}{$\Delta$, mm/s} & \multicolumn{2}{c}{$H_{hf}$, T} \\
 &  & before & after & before & after\\
\hline
FeBO$_3$ & 0.13(3) & 0.06(9) & 1.98(3) & 49.0(2) & 0 \\
Fe$_3$(Fe$_{1.766}$Si$_{0.234}$)(SiO$_4$)$_3$ & 0.19(2) & 0.25(6) & 0.89(1) & 0 & 0\\
CaFe$_2$O$_4$ & 0.16(3) at 5~K & 0.6 \& 1.2 & 1.1 \& 1.6 & 47.6 \& 45.4 at 5 K & 0\\
FeOOH & 0.3 at 6~K & --- & 2.5 at 50 K & 49 at 6~K  & 7.2 at 6~K\\
Fe$_2$O$_3$ & 0.22(3) & 0.71(2) & 0.76(3) & 50.15(6) & 0\\
\end{tabular}
\end{ruledtabular}
\end{table*}

All considered compounds demonstrate  similar behavior through spin transitions. The main features are \emph{(i)} drop in CS value, \emph{(ii)} disappearance of magnetic order at room temperature, and \emph{(iii)} LS state of ferric iron ion characterized by doublet with higher quadrupole splitting  (relative to HS values of Fe$^{3+}$). All data on hyperfine parameters of ferric iron in compounds of interest before and after spin transition are compiled in Table~\ref{MS_parameters}.

\subsubsection{\label{Isomer_chemical_shift}Isomer chemical shift}

The physical meaning of isomer shift in MS is an electron density on the probe nucleus which is almost fully created by $s$-electrons (with relativistic considerations there is also a minor $p$-electron contribution)~\cite{greenwood1971moessbauer}. One can write it in such form

\begin{equation}
\delta_{IS}=\frac{2\pi}{3}Ze^2[\langle r^2_{e}\rangle-\langle r^2_{g}\rangle]\Delta\psi^2(0)=\alpha\cdot\Delta\psi^2(0),
\end{equation}
where $Z$ is the proton number, $\langle r^2\rangle$ is the mean-squared nuclear radius of the excited and ground states, and $\Delta\psi^2(0)$ is the difference in electron density at the nucleus between the measured and reference compounds. The constant $\alpha$ is negative in the case of $^{57}$Fe.

The difference in the isomer shift between HS and LS states has been known since the early days of M\"{o}ssbauer spectroscopy~\cite{greenwood1971moessbauer}. As HS and LS ions have different radii, the metal-ligand bonds should be shorter for the LS configuration and, accordingly, an increased covalency of chemical bonds. This suggests that in the LS state, the occupancy of $4s$-orbitals increase which leads to higher electron density at the nucleus and to lower isomer shift. However, modern density functional theory (DFT) calculations have not found a correlation between the IS value and $4s$-L\"{o}wdin population~\cite{neese2002prediction}. In that paper the general inference was made that the metal-ligand bond length is important and that shorter bonds result in lower isomer shifts. In any event, bond length is a determinative factor and the IS value is necessarily lower for the LS iron ion.

In the compounds in the present study, the spin transition starts at CS values of $0.24\pm0.2$~mm/s at room temperature. As a reference value of isomer shift changes at spin transitions, we take the result for iron borate~---~0.13(3)~mm/s for the following reasons: (\emph{i}) the transition is isosymmetric (as distinct to the case of hematite), (\emph{ii}) the transition is first-order, i.e., abrupt (unlike in skiagite-iron-majorite, for example), and (\emph{iii}) the M\"{o}ssbauer spectra were measured at room temperature so the influence of changes in the second-order Doppler (SOD) shift should be minimal  (see discussion in Ref.~\onlinecite{menil1985systematic}).

The difference in CS values between HS and LS states is half the difference compared to coordination complexes ($\sim0.2$~mm/s)~\cite{nihei2007spin}. The smaller difference is probably related to smaller changes in the octahedral volume of Fe$^{3+}$O$_6$ (compare changes in bond length for different ligands in Ref.~\onlinecite{nihei2007spin}, Table~1). The relatively large discontinuity in values for FeOOH (Table~\ref{MS_parameters}) might be related to hydrogen bond symmetrization and corresponding significant changes in the SOD shift (in this case the zero-motion contribution may change significantly).%\footnote{moreover, the correctness of data processing is questionable as reported $H_{hf}$ (7.5~T) and $\Delta$ before magnetic ordering (2.5~mm/s) clearly indicate that usual perturbation treatment is not applicable and the problem requires solution of the Hamiltonian.}

\subsubsection{Second-order Doppler shift and center shift}

The second-order Doppler shift results in reduction of $\gamma$-ray energy due to relativistic time dilatation in the reference frame associated with the moving nucleus~~\cite{greenwood1971moessbauer}
\begin{equation}
 \label{SOD_shift}
\delta_{SOD}=-\frac{\langle v^2\rangle}{2c^2}E_{\gamma}=-\frac{\langle \epsilon_k\rangle}{Mc^2}E_{\gamma},
\end{equation}
where $E_{\gamma}$ is the $\gamma$-quantum energy, $\langle v^2\rangle$ is the mean-squared velocity of the nucleus in the crystal, $M$ is the nuclear mass, $c$ is the speed of light and $\langle \epsilon_k\rangle$ is the mean kinetic energy of the nucleus. Under the assumption that the nucleus is a harmonic oscillator, the mean kinetic energy is half of the vibrational internal energy and the SOD shift depends linearly on the internal energy of the nucleus. At the HS$\rightarrow$LS transition the magnitude of $\delta_{SOD}$ should increase as the decrease of the iron ionic radius causes a reduction of the polyhedron volume and, accordingly, leads to an increase of the force constants. Therefore, changes in $\delta_{SOD}$ will have the same sign as changes in the isomer shift.

Value of the SOD shift can be determined by means of nuclear inelastic scattering (NIS) data. From the NIS spectrum one can extract the partial phonon density of states (pDOS) of the M\"{o}ssbauer isotope and the vibrational internal energy can be directly determined by integrating the pDOS function~\cite{jones1973theoretical}

\begin{equation}\label{vibrational_energy}
  \epsilon_v=\frac{3}{2}\int_0^{\infty}E\coth\left(\frac{E}{2k_BT}\right){\mathcal D}(E)dE,
\end{equation}
where ${\mathcal D}(E)$ is the iron pDOS and $k_B$ is the Boltzmann constant.

For the compounds in this study, there is NIS data only for the skiagite--iron-majorite solid solution up to 56~GPa~\cite{Vasiukov2017sound}. As iron populates two different structural positions in this garnet, the corresponding pDOS is an average function of both. Nevertheless, these data can yield a reasonable estimation of changes in the SOD shift at the spin transition of Y-site Fe$^{3+}$ as $\delta_{CS}$ of X-site Fe$^{2+}$ does not show any evident anomalies in the vicinity of spin crossover (Fig.~\ref{Skiagite_moessbauer}a). There is a 4~meV increase of mean internal energy between 45 and 56~GPa~\cite{Vasiukov2017sound} that corresponds to 0.011(4)~mm/s change of $\delta_{SOD}$. This is about 10~\% of the observed variation in center shift upon HS-LS crossover.

One can also use siderite (FeCO$_3$) data~\cite{chariton_personal} to crosscheck this estimation. Siderite is isostructural with iron borate and contains carbon and ferrous iron instead of boron and ferric iron, respectively. Siderite also undergoes a HS$\rightarrow$LS transition at similar pressures ($\approx45$~GPa) and with similar changes in unit cell and octahedron volumes~\cite{lavina2010structure,merlini2013single}; hence the phonon properties of the crystal lattice should be similar to those of FeBO$_3$. Siderite NIS data show a 4~meV discontinuity at the spin transition~\cite{chariton_personal}, and the respective $\delta_{SOD}$ change is 0.011(2)~mm/s, similar to the estimation for the skiagite-iron-majorite solid solution.

The center shift extracted from M\"{o}ssbauer spectra is simply the sum of isomer and SOD shifts ($\delta_{CS}=\delta_{IS}+\delta_{SOD}$). Thus, our estimations suggest that reduction of $\delta_{CS}$ at the pressure-induced spin transition of Fe$^{3+}$ at ambient temperature is mainly caused by the isomer shift. The SOD contribution is not higher than 10~\%.

\subsubsection{Hyperfine magnetic field}

At the HS$\rightarrow$LS transition in octahedral coordination, the spin of ferric iron changes from 5/2 to 1/2. This means a huge drop of the iron magnetic moment, but Fe$^{3+}$ remains paramagnetic in the LS state as distinct from LS Fe$^{2+}$. Accordingly, the transition should cause a large drop of N{\'e}el (Curie) temperature. Indeed, in all studied compounds that were magnetically ordered in the HS state, the disappearance of magnetic ordering or a large decrease of the $H_{hf}$ value at the spin transition are observed.

Reliable experimental information about the $T_N$ pressure dependence is unfortunately absent. To estimate the magnetic critical temperature in the vicinity of the spin transition for the considered compounds, one can apply Bloch's law~\cite{bloch1966103}:
\begin{equation}\label{Blochlaw}
T_N\propto J \propto V^{\left(-10/3\right)},
\end{equation}
where $J$ is the exchange integral. For $V$ it is more reasonable to use the volume of the octahedron instead of the unit cell volume as it better determines the overlapping of electronic wave functions. The obtained estimations are listed in Table~\ref{Neel_temperatures}. Using the mean-field approximation and assuming that the exchange integral does not change substantially, one can obtain that the N{\'e}el temperature after the spin transition will be 3/35 of $T_N$ before the transition~\cite{gavriliuk2005high}. Although the constancy of the exchange integral through the spin transition is doubtful\footnote{for instance, through the spin transition the $\sigma$ hopping integral $t_{pd\sigma}$ for corner-shared octahedra will be replaced by the significantly smaller $\pi$ hopping $t_{pd\pi}$}, this result is satisfactory for rough estimations. The factor 3/35 means that existence of magnetic order at room temperature after the spin transition requires that the N{\'e}el temperature before the transition should be around 3500~K. Even the large $T_N$ in hematite (1675~K, Table~\ref{Neel_temperatures}) is substantially less than this value.

\begin{table}[t]
\caption{\label{Neel_temperatures}Estimated N{\'e}el temperatures from Bloch's law in the vicinity of the HS$\rightarrow$LS transition in the studied compounds.}
\begin{ruledtabular}
\begin{tabular}{cC{3cm}C{3cm}}
Compound & $T_N$ at ambient pressure,~K & $T_N$ just below transition,~K\\
\hline
FeBO$_3$ & 348 & 690\\
FeOOH & 393 & 649\\
CaFe$_2$O$_4$ & 200 & 370\\
Fe$_2$O$_3$ & 948 & 1675\\
\end{tabular}
\end{ruledtabular}
\end{table}

The hyperfine magnetic field results from interaction of the nuclear spin with its own electrons\footnote{there is also a magnetic field from dipolar interactions with magnetic moments of neighboring atoms but its value is usually less than 1~T and can be neglected} and can be  expressed as a sum of three contributions: Fermi contact interaction ($H_c$) and a dipolar interaction with orbital and spin momenta ($H_L$ and $H_S$, respectively) of the electrons\cite{greenwood1971moessbauer}:

\begin{equation}
\label{H_hf}
H_{hf}=H_c+H_L+H_S,
\end{equation}
\begin{equation}
H_c=\frac{8\pi}{3}g_e\mu_B\langle{\mathbf{S}}\rangle\sum_{ns}\left[\left|\psi_{ns}^{\uparrow}(0)\right|^2-\left|\psi_{ns}^{\downarrow}(0)\right|^2\right],
\end{equation}
\begin{equation}
H_L=g_e\mu_B\langle\frac{1}{r^3}\rangle\langle\mathbf{L}\rangle,
\end{equation}
\begin{equation}
H_S=g_e\mu_B\langle3\mathbf{r}(\mathbf{S}\cdot\mathbf{r})\frac{1}{r^5}-\mathbf{S}\frac{1}{r^3}\rangle,
\end{equation}
where $g_e$ is the electron spin $g$-factor, $\mu_B$ is the Bohr magneton, $\left|\psi_{ns}^{\uparrow\downarrow}(0)\right|^2$ is the electron density at the nucleus for a given $ns$ shell with spin parallel or antiparallel to the expectation value of the net electronic spin $\langle{\mathbf{S}}\rangle$, $\langle{\mathbf{L}}\rangle$ is the expectation value of orbital momentum and $r$ is the radial coordinate of electrons. As all terms in~(\ref{H_hf}) depend on $\langle{\mathbf{S}}\rangle$ or $\langle{\mathbf{L}}\rangle$, the corresponding $H_{hf}$ of Fe$^{3+}$ in HS and LS configurations should be significantly different.

The saturation $H_{hf}$ value of HS ferric iron in oxygen octahedra (electronic term $^6A_{1g}$) is usually around 50--55~T. For this electronic configuration orbital momentum is absent, the saturation value of $\langle{\mathbf{S}}\rangle$ is 5/2 and the observed hyperfine field arises mainly from the Fermi contact interaction term. In the LS state ($^2T_{2g}$ term) the saturated values of $\langle{\mathbf{S}}\rangle$ or $\langle{\mathbf{L}}\rangle$ can reach 1/2 and 1, respectively. So in this case $H_c$ should be around 11~T, but $H_L$ and $H_S$ can now also have comparable values. As $H_L$ and $H_S$ may have different signs relative to $H_c$, the resulting hyperfine fields may lie in the 0 to 20~T range.

There are only a scarce amount of data for $H_{hf}$ of Fe$^{3+}$O$_6$ in the LS state. Magnetic ordering was observed in FeOOH  with LS Fe$^{3+}$ that showed a hyperfine field of 7.2~T at 6~K (Table~\ref{MS_parameters}, but it is not clear if $H_{hf}$ was saturated) and magnetism was also reported in FeBO$_3$ below 50~K~\cite{gavriliuk2005high}. In the latter case perturbations of NFS spectra were detected, but it was not possible to fit them satisfactorily. Moreover, spin transitions in Fe$^{3+}$ can lead to metallization~\cite{ovchinnikov2008effect} with disappearance of magnetic moments (as in the case of CaFe$_2$O$_4$~\cite{greenberg2013mott}). It is clear that further systematic studies of magnetic ordering in phases containing high-pressure LS trivalent iron are required.

\subsubsection{Quadrupole splitting}

The main parameter of M\"{o}ssbauer spectra that is usually used to discriminate between HS and LS states is the quadrupole splitting. The strength of the quadrupole interaction depends on the electric field gradient created at the nucleus by the electron cloud (electronic contribution) and by neighboring ions (lattice contribution)~\cite{greenwood1971moessbauer}:
\begin{equation}
 q=(1-R)q_e+(1-\gamma_\infty)q_i,
\end{equation}
where $q$, $q_e$ and $q_i$ are the EFG (total, electronic and lattice contributions, respectively), and $R$ and $\gamma_\infty$ are the Sternheimer factors of shielding and antishielding, respectively.  Since the EFG is proportional to $r^{-3}$, firstly the electronic part generally is the dominant contribution to the EFG and secondly, in the case when the lattice contribution plays a principle role the EFG will be mainly dominated by the first coordination sphere (since the second neighbors at $2r$ distance produce an EFG that is 8 times weaker). The above-mentioned correlation between decrease of the iron octahedron distortion and quadrupole shift in iron borate is a good example of the latter case.

As fully-filled or half-filled $e_g$ and $t_{2g}$ orbitals do not produce an EFG (see Table 4.2 in Ref.~\onlinecite{guetlich2010moessbauer}), the quadrupole splitting in the case of HS Fe$^{3+}$ is mainly related to the lattice contribution and is generally small. In octahedrally coordinated Fe$^{3+}$ in the LS state, the electron term is $^2T_{2g}$ with one unpaired electron on the $t_{2g}$ level and, therefore, for ferric iron at the spin transition the main contribution to the EFG changes from lattice to electronic (for ferrous iron the situation is opposite). This should cause a significant increase of the QS value and, indeed, this tendency is observed in all compounds with the exception of Fe$_2$O$_3$ (Table~\ref{MS_parameters}) where the polyhedron type may change.

There is a large spread of QS values of the Fe$^{3+}$ LS state: from 0.9~mm/s in CaFe$_2$O$_4$ and\linebreak skiagite-iron-majorite to 2.5~mm/s in FeOOH (at 50~K). Note that the lowest value of $\Delta$ is significantly smaller than the corresponding value 1.9~mm/s reported for LS Fe$^{3+}$ in  coordination complexes~\cite{nihei2007spin}. The analysis of the EFG in LS Fe$^{3+}$ is  analogous to the approach for HS Fe$^{2+}$~\cite{ingalls1964electric}. The additional electron on the $t_{2g}$ level is replaced by a hole and the EFG will be controlled by $t_{2g}$ manifold splitting and temperature. This explains the spread of QS values in the LS state through the different degrees of octahedral distortion in the studied compounds. The temperature dependence of $\Delta$ for LS Fe$^{3+}$ has been observed for FeOOH and CaFe$_2$O$_4$~\cite{xu2013pressure,greenberg2013mott}.

\subsection{Structure data}

\subsubsection{\label{Oct_volume_section}Volume of Fe$^{3+}$O$_6$ octahedron at spin transition}

\begin{figure}[h]
\includegraphics[width=8.6cm, keepaspectratio=true]{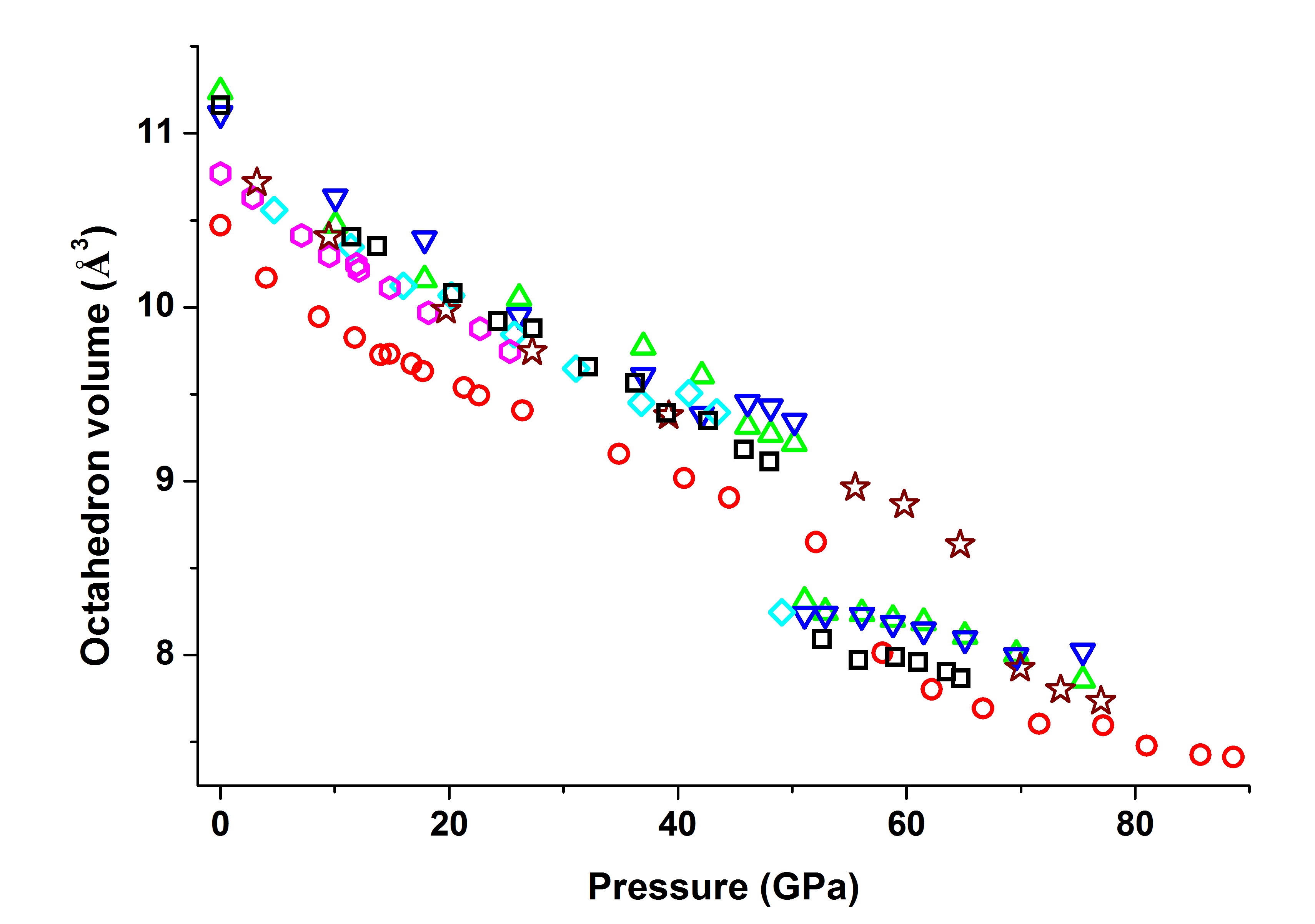}
\caption{\small{\label{Oct_volumes_vs_pressure}Volume of Fe$^{3+}$O$_6$ octahedra as a function of pressure for different compounds. Black squares correspond to iron borate, red circles to skiagite, green and blue triangles to two different structural positions in calcium ferrite, cyan diamonds to goethite, magenta hexagons to hematite and brown stars to andradite.}}
\end{figure}

Fig.~\ref{Oct_volumes_vs_pressure} shows the dependence of the octahedron volumes in the studied compounds as a function of pressure. One can see that the spin transition of Fe$^{3+}$ starts in the pressure range 45--60~GPa and over a remarkably narrow range of octahedron volume --- 8.9--9.3~{\AA}$^3$. This suggests that the spin transition is controlled by the electronic density inside the octahedron and explains why the spin transition starts within a narrow range of CS values (see section~\ref{Isomer_chemical_shift}).

\begin{table*}[t]
\caption{\label{CF_estimation}Crystal field parameters and the estimated polyhedron volumes at transition onset.}
\begin{ruledtabular}
\begin{tabular}{ccccccc}
Compound & $10D_q$~(cm$^{-1}$) & B~(cm$^{-1}$) & $\beta$ & $V_0$~({\AA}$^3$) & $V_t$ from eq.~(\ref{Polyh_volume})~({\AA}$^3$) & exp. $V_t$~({\AA}$^3$)\\
\hline
FeBO$_3$ & 12700~\cite{malakhovskii1976optical} &  680 & 0.68 & 11.61 & 9.0 & 9.1\\
Fe$_2$O$_3$ & 14000~\cite{sherman1985electronic} &  540 & 0.54 & 10.77 & 10.2 & 9.0\\
FeOOH & 15320~\cite{sherman1985electronic} & 590 & 0.59 & 10.81 & 10.3 & 9.3\\
Ca$_3$Fe$_2$Si$_3$O$_{12}$ & 12600~\cite{burns1993mineralogical} & 593 & 0.59 & 10.9 & 9.2 & 8.9\\
%\hline
%FeCO$_3$  & 10325~\cite{lobanov2015optical} & 747 & 0.84 & 13.2 & 10.6 & 10.3 \\
%\hline
\end{tabular}
\end{ruledtabular}
\end{table*}

In crystal field theory the crystal field splitting parameter determines the strength of the electrostatic potential created by anions and is directly related to the distance between cation and anions ($D_q\propto r^{-5}$ for an ideal octahedron, ch.~2 in Ref.~\onlinecite{figgis2000ligand}). Using Tanabe-Sugano diagrams~\cite{tanabe1954absorptionI,*tanabe1954absorptionII} one can estimate the transition bond length at which the spin transition should happen. So, for the polyhedron volume we can write:
\begin{equation}
V_t=V_0\left(\frac{D_q^0}{D_q^t}\right)^{3/5},
\label{Polyh_volume}
\end{equation}
where $V$ is the octahedron volume, and the indices 0 and $t$ correspond to ambient pressure and at the spin transition, respectively. The results of this estimation for the compounds under consideration are collected in Table~\ref{CF_estimation}.  One can see that this simplest model overestimates the transition volume notably in the case of hematite and goethite. If predictions would be accurate, the spin transition in these compounds would take place at 20--25 GPa. Note that this discrepancy is not related to covalent effects. The degree of covalency can be estimated using the nephelauxetic ratio $\beta$ (smaller values of $\beta$ indicate more covalent bonds, while unity means a purely ionic bond). Despite the fact that hematite and goethite are more covalent than iron borate, Eq.~(\ref{Polyh_volume}) predicts nearly the correct octahedron volume for andradite, which has the same nephelauxetic ratio as hematite (Table~\ref{CF_estimation}).

\begin{table}[h]
\caption{\label{HS_EoS_octahedron}Equation of state parameters of the HS ferric iron octahedron in the studied compounds. The Birch-Murnagham (BM) EoS of both 2nd and 3rd order was used. For calcium ferrite, the average volume of two octahedra (for iron in two distinct crystallographic positions) was used.}
\begin{ruledtabular}
\begin{tabular}{ccccc}
Compound & EoS order & $V_0$ ({\AA}$^3$) & $K_0$ (GPa) & $K'_0$\\
\hline
FeBO$_3$ & 3rd & 11.16(fixed) & 166(9) & 3.8(6)\\
   & 2nd & 11.14(5) & 166(8) & 4\\
CaFe$_2$O$_4$ & 3rd & 11.18 (fixed) & 170(10) & 5.2(8)\\
& 2nd & 11.18(fixed) & 186(2) & 4\\
Ca$_3$Fe$_2$(SiO$_4$)$_3$ & 3rd & 10.88(1) & 202(4) & 3.62(16)\\
& 2nd & 10.89(1) & 194(2) & 4\\
Fe$_2$O$_3$ & 3rd & 10.77(1) & 200(20) & 5(3)\\
& 2nd & 10.77(1) & 209(6) & 4\\
FeOOH & 3rd & 10.75(fixed) & 202(11) & 5(1)\\
& 2nd & 10.75(fixed) & 214(4) & 4\\
\end{tabular}
\end{ruledtabular}
\end{table}

This problem is most likely related to octahedral distortions. In both hematite and goethite, octahedral Fe-O bonds divide into two groups (three bonds in each) with different lengths:  approximately 1.95 and 2.10~{\AA} at ambient conditions. Electrostatic potentials with symmetry lower than cubic contain additional terms that have  different power dependencies on $r$ (for instance, the term $\propto r^{-3}$ for a trigonally distorted octahedron, see ch.~2 in Ref.~\onlinecite{figgis2000ligand}). Therefore, consideration of the proper electrostatic potentials for hematite and goethite should include a correct estimation of the transition volumes, but such an examination is beyond the scope of this work.

%Note that as trivalent iron $D_q$ can be up to 40~\% larger than for divalent iron at ambient conditions~\cite{gutlich2013spin}, at first sight, one would expect higher transition pressure for compounds with ferrous iron. However, as more ionic bonds of Fe$^{2+}$ are substantially more compressible, the transition pressure close to ferric iron~\cite{badro2003iron,lavina2010structure}.

In all studied compounds, iron octahedra show very similar compressibility (Fig.~\ref{Oct_volumes_vs_pressure}).\linebreak Skiagite-iron-majorite solid solution data deviate from the rest of the examined compounds due to the mixed population of the Y-site by iron and silicon, because the SiO$_6$ octahedron is significantly smaller than one for Fe$^{3+}$O$_6$ (for example, in the ilmenite- and perovskite-type MgSiO$_3$ polymorphs the volume is around 7.64~\AA$^3$ at ambient conditions~\cite{horiuchi1982ilmenite_type,horiuchi1987perovskite_type}). Owing to this, XRD provides an average picture with a reduced volume of the (Fe,Si)O$_6$ octahedron (Fig.~\ref{Oct_volumes_vs_pressure}). At spin crossover the volumes of iron octahedra decrease and approach those of silicon octahedra, where the difference between the average volume of the (Fe,Si)O$_6$ octahedra and the rest of the data becomes small above 60~GPa (Fig.~\ref{Oct_volumes_vs_pressure}).

%Supposing the independence of octahedron volume on the specific surroundings (assumption analogous to Vegard's law), we can estimate volume of the mixed octahedron. On the basis of studied compounds we can assume 10.9~\AA$^3$ value for iron octahedron at ambient conditions (see Fig.~\ref{Oct_volumes_vs_pressure}). As the octahedron contains 23.4\% of Si~\cite{ismailova2015high} the calculated average octahedron volume is 10.52~\AA$^3$. This value is in perfect agreement with experimental value 10.47~\AA$^3$ (Fig.~\ref{Oct_volumes_vs_pressure}).

The isothermal bulk moduli of Fe$^{3+}$O$_6$ octahedra in different compounds at ambient conditions vary between 170 and 210~GPa (Table~\ref{HS_EoS_octahedron}). The $K_0$ value correlates with the octahedron volume at ambient pressure and, apparently, it is related to the length of Fe-O bonds (the shorter the bond, the more incompressible the octahedron). In the vicinity of the spin transition the average value of $K$ is around 400 GPa (iron borate has the lowest value of 340(20) GPa). Currently, only iron borate and calcium ferrite have sufficient data points above the spin transition for estimation of the LS octahedron EoS~\cite{Greenberg_be_published,merlini2010letter}. For FeBO$_3$ the parameters of the EoS (BM, 2nd order) are $V_0=9.6(4)$~\AA$^3$ and $K_0=225(65)$~GPa, and for CaFe$_2$O$_4$ they are $V_0=9.4(2)$~\AA$^3$ and $K_0=320(50)$~GPa. As LS ferric iron has a significantly smaller ionic radius, one can expect that elastic moduli should increase at the HS$\rightarrow$LS transition. Indeed, an EoS comparison of HS and LS Fe$^{3+}$O$_6$ octahedra shows that at 50~GPa, the bulk modulus changes from 340(20)~GPa to $410(70)$~GPa and from $400(20)$~GPa to $510(50)$~GPa %(3rd order EoS was used, see Table~\ref{HS_EoS_octahedron})
for FeBO$_3$ and CaFe$_2$O$_4$, respectively.

\subsubsection{Cooperativity at spin transition}

Available single-crystal XRD data show that pressure-induced spin transitions in inorganic compounds have a strong tendency to result in isosymmetric transitions. From the phenomenological theory of phase transitions, it follows that they can proceed either as first-order transitions or, beyond the critical point, a crossover in which there is no discontinuity in any free-energy derivative~\cite{christy1995isosymmetric}.

The abundance ratio of different spin states would be controlled by Boltzmann factor if cations change spin states independently. However, since cations in different spin states have different ionic radii, a spin transition of one ion introduces a strain field to the crystal lattice (one can consider the ion in a different spin state as an impurity). The coupling with this strain field determines the strength of cooperative behavior at the spin transition and, accordingly, the critical temperature ($T_{cp}$) at which the first-order phase transition is changed by crossover behavior. In general, the elastic interaction between impurities in a crystal is complex and may even lead to superstructures with cations coexisting in different spin states~\cite{khomskii2004superstructures}. However, to our knowledge, no single-crystal XRD study of pressure-induced spin transition has observed such superstructures. We therefore limit our discussion to simple isosymmetric transitions, as they involve all principal features of the spin-transition phenomenon.

%For the elastically isotropic medium the coupling is infinite-range attractive interaction~\cite{eshelby1956continuum}. However, in reality the situation can be much more complicated. The crystal is inherently anisotropic medium and for anisotropic elastic interaction the directional dependence of interaction between different spin states may appear (note, that interactions can be also \emph{repulsive} is some directions) and as result, it may lead to formation of superstructures with ions in different spin states~\cite{khomskii2004superstructures}.

In the crossover regime, there is \emph{dynamical spin state equilibrium} at the spin transition, so cations permanently change their spin state. The rate of these changes ($t_s$) is an important parameter, since different experimental techniques have different characteristic measurement times ($t_m$, $10^{-7}$~s for $^{57}$Fe MS). Hence, depending on the ratio $t_s/t_m$, one technique can show an averaged spin state while another can show the \emph{apparent} static coexistence of spin states.  Since phase coexistence is an inherent feature of first-order phase transitions, it is not necessarily a trivial task to determine which mechanism of spin transition is observed, which may lead to wrong conclusions. In this respect diffraction techniques have an advantage, because they average over ensemble and not time period. Hence if a spin transition proceeds in a first-order manner, one will see peak splitting corresponding to spin-domains with different volumes (see siderite example in Ref.~\onlinecite{lavina2010structure}), while for crossover it will be simply a single phase with averaged volume. In addition, results of powder DAC experiments should be considered with caution because of the tendency to broaden the transition region (see comparison of powder and single-crystal experiments in Ref.~\onlinecite{cerantola2015high}).

For the examined compounds (Table~\ref{Used data}) the condition of cooperativity is as follows: if iron octahedra share common oxygen atoms and form an infinite framework, the compound shows strong cooperative behavior at pressure-induced spin transitions (by ``strong'' we mean that the critical point lies above room temperature). If this condition is fulfilled there are favorable conditions for magnetic ordering by means of superexchange interactions. Indeed, all such studied compounds in the HS state are magnetically ordered at room temperature in the vicinity of the spin transition (see Table~\ref{Neel_temperatures}). However, if the compound undergoes an isosymmetric (according to XRD) HS$\rightarrow$LS transition and magnetic ordering disappears in the LS state, when, strictly speaking, the transition is not isosymmetric, because at such transition the symmetry relative to time reversal is changed. This leads to important consequences: (\emph{i}) the spin transition in magnetically-ordered compounds can only proceed as a first-order phase transition, and (\emph{ii}) the temperature of the critical point cannot be lower than the N{\'e}el (Curie) temperature of the HS phase in the vicinity of the spin transition. These conclusions are valid if the spin transition is the driving force of the transition.

Among the studied compounds, only the spin transitions in the Y-site of the garnet structure (andradite and skiagite--iron-majorite solid solution) show explicit crossover behavior at room temperature within $\sim10$~GPa. This is in agreement with the formulated cooperativity condition, since Y-site octahedra do not share oxygen atoms (Fig.~\ref{Sk_Mj_structure}a). Note that this is a clear structure feature, where the spin transition of Mn$^{3+}$ ($d^4$ configuration) in the octahedral site of the hydrogarnet henritermierite proceeds in the same way~\cite{friedrich2015pressure}.

An interesting transition takes place in yttrium iron garnet (YIG) Y$_3$Fe$_5$O$_{12}$ in which ferric iron populates both the Y-sites and tetrahedral Z-sites. Although YIG is isostructural to the garnets discussed above, it undergoes a different phase transition. At pressures characteristic for the Fe$^{3+}$O$_6$ spin transition ($50\pm1$~GPa), there is abrupt irreversible amorphization of this garnet~\cite{gavriliuk2006equation,stan2015high} that is not related to the mechanical instability~\cite{stan2015high}. YIG remains ferrimagnetic ($T_C=559$~K at ambient conditions) up to amorphization, so a spin transition according to our phenomenological arguments can proceed only in a first-order manner.

The Fe$^{3+}$O$_4$ tetrahedron is a ``metastable'' polyhedron at high pressure due to the relatively large ionic radius of ferric iron (49 pm \emph{vs} 26 pm of Si$^{4+}$~\cite{shannon1976revised}) and the general tendency to increase coordination number with pressure~\cite{hazen1982comparative}.  Indeed, XRD and M\"{o}ssbauer data of amorphous YIG show evidence that iron has six-fold coordination in this state~\cite{gavriliuk2006equation,[][; however Fe$^{3+}$ likely remains in the HS state as there is no drop in CS value for octahedrally-coordinated iron and on decompression the hyperfine parameters do not show any significant changes.]lyubutin2005magnetic}. Moreover, YIG is metastable at room temperature above 31~GPa and transforms to the perovskite structure (in which the minimum cation coordination number is six) upon heating~\cite{stan2015high}, which is typical behavior for members of the garnet family. There is a large kinetic barrier between the garnet and perovskite structures, which, to our knowledge, cannot be overcome by any garnet at room temperature. Therefore, it is the instability of HS Fe$^{3+}$O$_6$ octahedra upon reaching the critical volume that serves as a trigger for this amorphization, allowing the kinetic barrier to be  overcome (partially in this case). The cooperative behavior of iron octahedra might be crucial for the process. Indeed, spin transition by crossover does not lead to such consequences for the garnet examples above (although they are also certainly metastable at spin crossover pressures). The appearance of $\iota$-Fe$_2$O$_3$ at the transition of $\alpha$-Fe$_2$O$_3$ to $\zeta$-Fe$_2$O$_3$ (see section~\ref{hematite_results}) might also be related to this phenomenon.

In the case of solid solutions, the highest $T_{cp}$ value will be seen in the iron end-member. Reduction of the amount of iron in the system will decrease the cooperativity of iron octahedra and, accordingly, the critical point temperature. In natural systems pressure-induced spin transitions are inferred to occur in the Earth and exoplanet interiors. In view of the above considerations the iron spin transition in the Earth's mantle should tend to crossover behavior for the following reasons: (\emph{i})~mantle minerals are solid solutions with relatively low concentrations of iron that promote a decrease of the cooperativity of iron ions, and (\emph{ii}) the temperatures of the mantle are high ($>2000$~K) at spin-transition pressures.

\subsection{Center shift \emph{vs} polyhedron volume}

\begin{figure}[h]
\includegraphics[width=8.6cm, keepaspectratio=true]{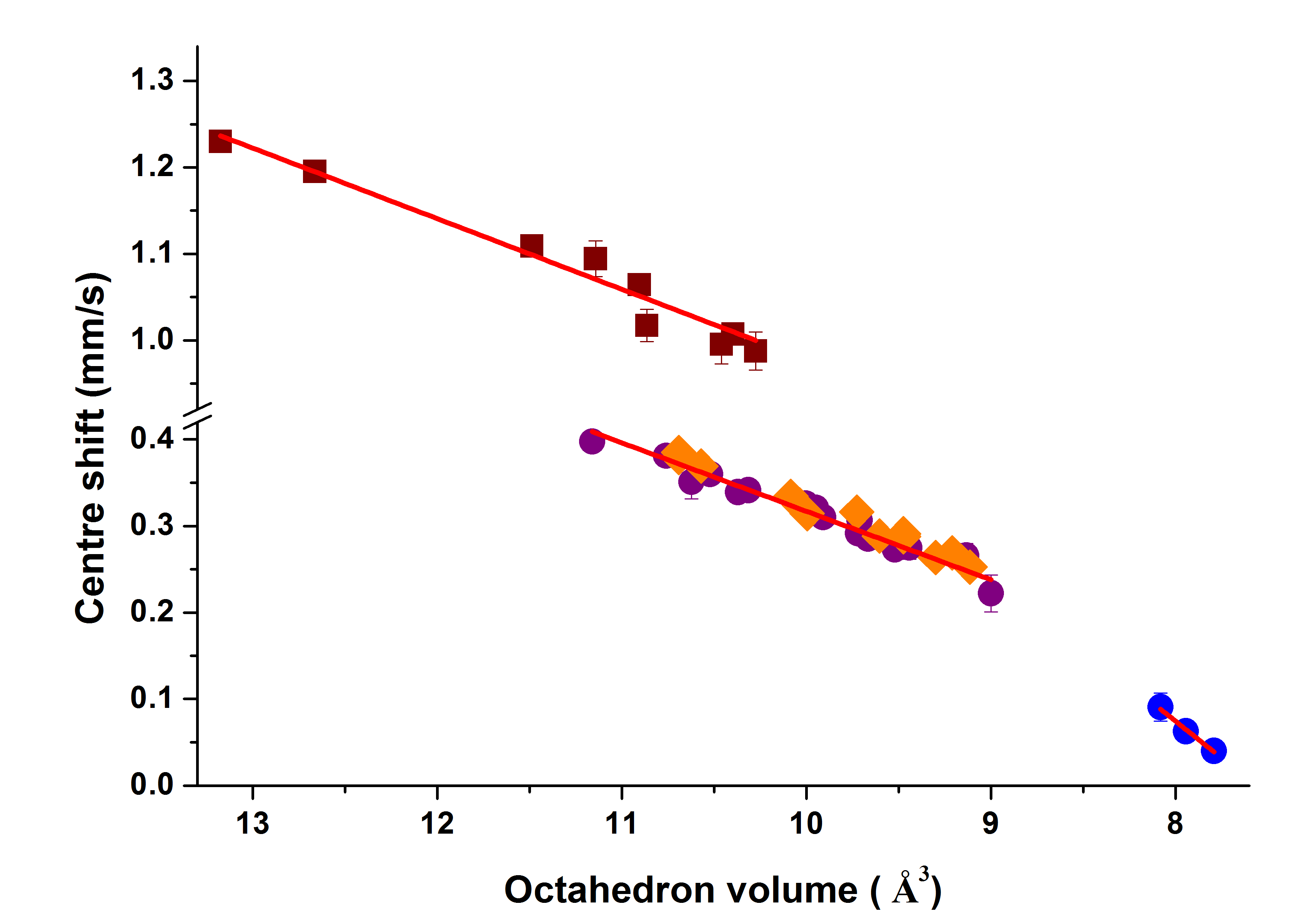}
\caption{\label{CS_oct_volume}Center shift as a function of the octahedron volume. Within experimental uncertainties data show a linear dependence. Purple and blue circles are data of FeBO$_3$ (HS and LS, respectively), orange diamonds correspond to $\alpha$-Fe$_2$O$_3$, and brown squares belong to FeCO$_3$ (single-crystal data from Ref.~\onlinecite{cerantola2015high}). The red lines are linear fits to the underlying points (the middle line is a fit of the FeBO$_3$ HS points). In the HS state, slope values are the same for both ferric (FeBO$_3$, $\alpha$-Fe$_2$O$_3$) and ferrous (FeCO$_3$) iron ions. The LS state in iron borate is more sensitive to changes of octahedron volume.}
\end{figure}

\begin{figure*}[t]
\includegraphics[width=17.8cm, keepaspectratio=true]{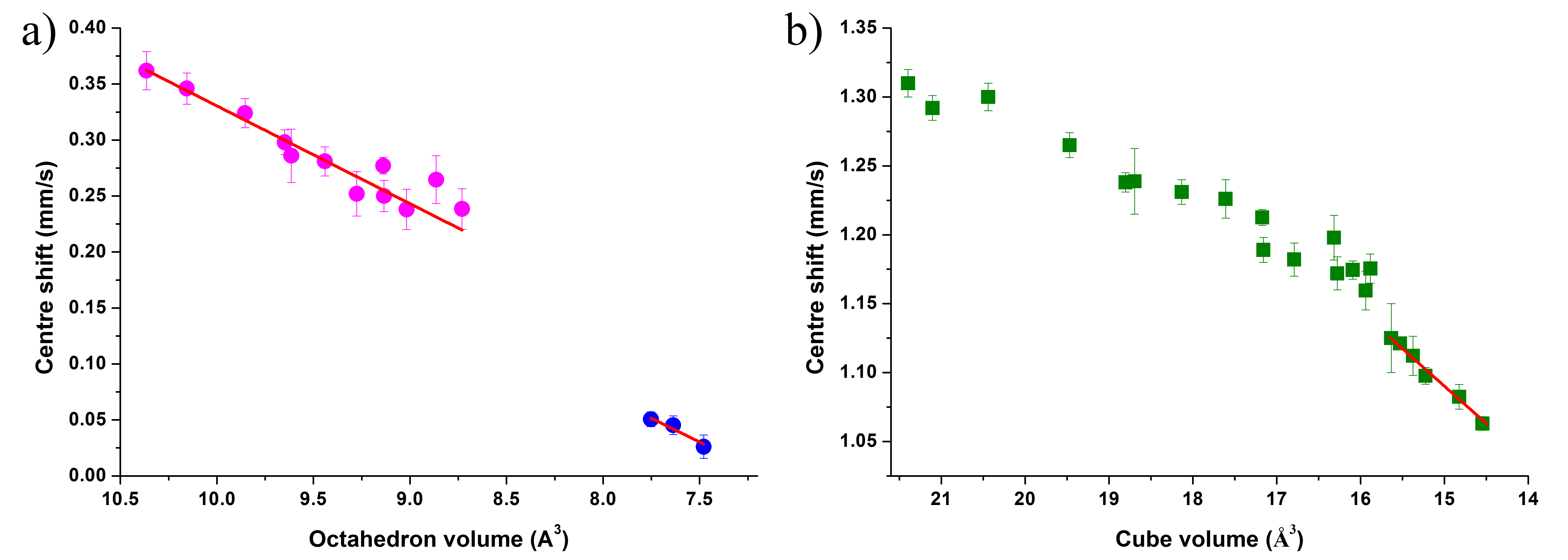}
\caption{\label{CS_pol_volume_skiagite}Volume dependence of CS in skiagite-iron-majorite solid solution. a) CS of Y-site HS Fe$^{3+}$ (pink circles) also depends linearly on octahedron volume, but LS data (blue circles) show almost the same slope in contrast to FeBO$_3$.  b) CS of X-site Fe$^{2+}$ (in distorted cube) has a much weaker non-linear dependence. The red lines are linear fits to the underlying points.}
\end{figure*}

Single-crystal XRD experiments provide a unique opportunity to study the dependence of CS on the polyhedron volume (Figs.~\ref{CS_oct_volume} and~\ref{CS_pol_volume_skiagite}). %Till now it was done only for metals and alloys with highly symmetric structures~\cite{williamson1978influence} where atomic coordinates are totally fixed by symmetry.
In addition to the compounds measured in this work, we also used for comparison single-crystal M\"{o}ssbauer data of siderite~\cite{cerantola2015high}. Using structure data from Ref.~\onlinecite{lavina2010structure} we obtained the following 2nd BM EoS parameters for the Fe$^{2+}$O$_6$ octahedron: $V_0=13.17(4)$~\AA$^3$ and $K_0=108(3)$~GPa. Hence the octahedron of ferrous iron is larger by $\sim2$~\AA$^3$ at ambient conditions (see Fig.~\ref{Oct_volumes_vs_pressure}) and more compressible relative to ferric iron octahedra.

In all studied compounds the CS value varies linearly with octahedron volume. It is interesting that for the HS octahedra, linear fits have the same slope (Fig.~\ref{CS_oct_volume}) within uncertainty, regardless of valence state: 0.079(4), 0.079(5) and 0.082(4)~mm/s$\cdot${\AA}$^{-3}$ for FeBO$_3$, $\alpha$-Fe$_2$O$_3$ and FeCO$_3$, respectively. Y-site Fe$^{3+}$ in skiagite-iron-majorite solid solution shows the same linear dependence in the HS state (Fig.~\ref{CS_oct_volume}a) with slope 0.087(6)~mm/s$\cdot${\AA}$^{-3}$, but one should note the systematic error in this value because the average (Fe,Si)O$_6$ volume was used.

This similarity of slopes suggests that the volume dependence of CS is governed by the same mechanism in all investigated compounds. The noted difference in pressure dependence of isomer shift between Fe$^{2+}$ and Fe$^{3+}$ compounds~\cite{williamson1978influence} is therefore simply related to the higher compressibility of ferrous iron (at least in the case of octahedral coordination). High-pressure NIS data of siderite~\cite{chariton_personal} show that $\delta_{SOD}$ also depends linearly on octahedron volume with 0.0050(1)~mm/s$\cdot${\AA}$^{-3}$ coefficient, which is 6~\% of the observed volume dependence of $\delta_{CS}$.

%Data from [Lin2011] show that in $\alpha$-Fe$_2$O$_3$ the SOD contribution is not higher than 8~\%.

The data clearly show that the CS discontinuity between two spin states is caused not only by the volume difference, but also to changes in the electronic configuration, since the LS values do not lie on continuations of HS lines (Fig.~\ref{CS_oct_volume} and~\ref{CS_pol_volume_skiagite}a). In the LS state $\delta_{CS}$ is more sensitive to volume changes in the FeBO$_3$ case (Fig.~\ref{CS_oct_volume}), while LS Fe$^{3+}$ in skiagite-iron-majorite shows roughly the same slope as the HS state (Fig.~\ref{CS_pol_volume_skiagite}a). The number of data points is too small, however, for definite conclusions. Also note that in the latter case the CS values can be affected by strong overlap with singlet component (Fig.~\ref{Sk_Mj_structure}b).

The volume dependence of $\delta_{CS}$ for different types of coordination polyhedron can be obtained from the skiagite-iron-majorite data.  The CS variation of cubic X-site Fe$^{2+}$ is nonlinear and considerably weaker than for the case of the octahedron (Fig.~\ref{CS_pol_volume_skiagite}b). A linear fit of the data below 16~\AA$^3$ (pressure above 60~GPa) has a slope of 0.054(3)~mm/s$\cdot${\AA}$^{-3}$.

\section{Conclusions}

%At ambient pressure Fe$^{3+}$ in the octahedral oxygen environment always has a high spin state ($^6A_{1g}$).

We performed a comparative study of the spin transition in Fe$^{3+}$O$_6$ octahedra in a number of different compounds using MS in combination with single-crystal XRD data. Analysis of the obtained data shows that the most universal and unambiguous evidence of the HS$\rightarrow$LS transition among hyperfine parameters is a drop of the center shift value ($\geq0.13$~mm/s). One significant advantage is the temperature independence of the drop, which is not the case for quadrupole splitting, since for the studied case the QS values for the LS state can be close to HS values before spin transition. However, the temperature dependence of $\Delta$ within the LS state can be a reliable indicator.

We argue that $\zeta$-Fe$_2$O$_3$ is an intermediate phase in the reconstructive phase transition between $\iota$-Fe$_2$O$_3$ and $\theta$-Fe$_2$O$_3$. The interpretation of M\"{o}ssbauer data in the framework of the existing model for the structure of $\zeta$-Fe$_2$O$_3$~\cite{bykova2013novel} is not consistent with observations. Based on behavior of hyperfine parameters in the Fe$_2$O$_3$ system, we conclude a coexistence of HS and LS Fe$^{3+}$ in $\zeta$-Fe$_2$O$_3$ with LS Fe$^{3+}$ occupying trigonal prisms.

Structural data reveal that for all studied compounds, the spin transition starts within a narrow range of octahedral volumes: 8.9--9.3 \AA$^3$. Taking into account the compressibility of Fe$^{3+}$O$_6$ octahedra, this corresponds to the 45--60 GPa pressure range. The simple ideal octahedral model from crystal field theory predicts transition volumes with reasonable accuracy, but highly distorted octahedra require a more elaborate approach.

Spin transitions usually lead to isosymmetric structural transitions. Two scenarios are possible: supercritical, crossover behavior or first-order phase transition. The degree of cooperativity in the behavior of iron octahedra (controlled by elastic interactions between ions in different spin states) plays a crucial role, determining the position of the critical point on the phase diagram. From phenomenological arguments, it follows that in magnetically-ordered compounds the spin transition can proceed only in a first-order manner. Moreover, experiments show that cooperative behavior is preserved at least at room temperature if iron octahedra share common oxygen atoms. We argue that instability of iron octahedra on reaching the critical volume, together with cooperative behavior, is important for metastable phases as it may promote the overcoming of kinetic barriers, even at low temperatures. We note that these conclusions are applicable in general to spin transition phenomenon, regardless of the specific transition ion.

We have demonstrated that the center shift of HS iron depends linearly on octahedral volume with the same slope, regardless of oxidation state (0.08~mm/s$\cdot${\AA}$^{-3}$). NIS data show that the SOD contribution to this dependence is less than 10~\%. Such data can be used for determination of the isomer shift calibration constant. The center shift of ferrous iron in cubic coordination is less sensitive to the volume changes and the dependence is nonlinear.

\begin{acknowledgments}
Authors are grateful to Prof. V. Dmitriev and Dr. D. Chernyshov for fruitful discussion and Dr. E. Greenberg for provided data. We thank the European Synchrotron Radiation Facility for provision of synchrotron radiation (ID18). N.D. thanks the German Research Foundation (Deutsche Forschungsgemeinschaft, DFG, projects no. DU 954-8/1 and DU 954-11/1) and the Federal Ministry of Education and Research, Germany (BMBF, grants no. 5K13WC3 and 5K16WC1) for financial support. L.D. acknowledges DFG funding through projects the CarboPaT Research Unit FOR2125.
\end{acknowledgments}

\bibliography{Bibliography/My_bib}

\end{document}